\documentclass[twocolumn,amsmath,amssymb,floatfix,prl,footinbib,superscriptaddress]{revtex4-2}

\usepackage{amsmath,amssymb,natbib,bm,graphicx,url,epsf,color}
\usepackage[english]{babel}
\usepackage{wasysym}
\usepackage{MnSymbol}
\usepackage{amsmath}
\usepackage{graphicx}
\usepackage{pdfpages}
\usepackage[colorinlistoftodos]{todonotes}
\usepackage[colorlinks=true, allcolors=blue]{hyperref}

\newcommand{\QH}{QH~}

\newcommand{\Vdc}{V_{\mathrm{DC}}}

\newcommand{\kB}{k_{\mathrm{B}}}

\makeatletter
\AtBeginDocument{\let\LS@rot\@undefined}
\makeatother

\makeatletter
\def\NAT@spacechar{}
\makeatother

\begin{document}
\title{Enhanced shot noise in graphene quantum point contacts with electrostatic reconstruction}

\author{M. Garg}\thanks{Equal contribution.}
\affiliation{Universit\'e Paris-Saclay, CEA, CNRS, SPEC, 91191 Gif-sur-Yvette cedex, France
}
\affiliation{Department of Physics, Indian Institute of Technology Roorkee, Uttarakhand 247667, India
}

\author{O. Maillet}\thanks{Equal contribution. Contact: olivier.maillet@cea.fr}
\affiliation{Universit\'e Paris-Saclay, CEA, CNRS, SPEC, 91191 Gif-sur-Yvette cedex, France
}

\author{N. L. Samuelson}
\affiliation{Department of Physics, University of California at Santa Barbara, Santa Barbara CA 93106, USA}

\author{T. Wang}
\affiliation{Department of Physics, University of California at Berkeley, Berkeley, CA 94720, USA}
\affiliation{Material Science Division, Lawrence Berkeley National Laboratory, Berkeley, CA 94720,
USA}

\author{J. Feng}
\affiliation{Graduate Group in Applied Science \& Technology, University of California, Berkeley, California 94720, USA}

\author{L. A. Cohen}
\affiliation{Department of Physics, University of California at Santa Barbara, Santa Barbara CA 93106, USA}

\author{A. Zhang}
\affiliation{Universit\'e Paris-Saclay, CEA, CNRS, SPEC, 91191 Gif-sur-Yvette cedex, France
}

\author{K. Watanabe}
\affiliation{Research Center for Electronic and Optical Materials, National Institute for Materials Science, 1-1 Namiki, Tsukuba 305-0044, Japan
}
\author{T. Taniguchi}
\affiliation{Research Center for Materials Nanoarchitectonics, National Institute for Materials Science,  1-1 Namiki, Tsukuba 305-0044, Japan
}

\author{P. Roulleau}\thanks{Contact: preden.roulleau@cea.fr}
\affiliation{Universit\'e Paris-Saclay, CEA, CNRS, SPEC, 91191 Gif-sur-Yvette cedex, France
}

\author{M. Sassetti}
\affiliation{
Dipartimento di Fisica, Universit\`a di Genova, Via Dodecaneso 33, 16146, Genova, Italy
}
\affiliation{
SPIN-CNR, Via Dodecaneso 33, 16146, Genova, Italy
}

\author{M. Zaletel}
\affiliation{Department of Physics, University of California at Berkeley, Berkeley, CA 94720, USA}
\affiliation{Material Science Division, Lawrence Berkeley National Laboratory, Berkeley, CA 94720,
USA}

\author{A. F. Young}
\affiliation{Department of Physics, University of California at Santa Barbara, Santa Barbara CA 93106, USA}

\author{D. Ferraro}
\affiliation{
Dipartimento di Fisica, Universit\`a di Genova, Via Dodecaneso 33, 16146, Genova, Italy
}
\affiliation{
SPIN-CNR, Via Dodecaneso 33, 16146, Genova, Italy
}

\author{P. Roche}
\affiliation{Universit\'e Paris-Saclay, CEA, CNRS, SPEC, 91191 Gif-sur-Yvette cedex, France
}

\author{F.D. Parmentier}\thanks{Contact: francois.parmentier@phys.ens.fr}
\affiliation{Universit\'e Paris-Saclay, CEA, CNRS, SPEC, 91191 Gif-sur-Yvette cedex, France
}
\affiliation{Laboratoire de Physique de l’Ecole normale sup\'erieure, ENS, Universit\'e PSL,
CNRS, Sorbonne Universit\'e, Universit\'e Paris Cit\'e, F-75005 Paris, France
}

\date{\today}

\begin{abstract}
 Shot noise measurements in quantum point contacts are a powerful tool to investigate charge transport in the integer and fractional quantum Hall regime, in particular to unveil the charge, quantum statistics and tunneling dynamics of edge excitations. In this letter, we describe shot noise measurements in a graphene quantum point contact in the quantum Hall regime. At large magnetic field, the competition between confinement and electronic interactions gives rise to a quantum dot located at the saddle point of the quantum point contact. We show that the presence of this quantum dot leads to a $50-100~\%$ increase in the shot noise, which we attribute to correlated charge tunneling. Our results highlight the role played by the electrostatic environment in those graphene devices.
\end{abstract}

\maketitle

Quantum point contacts (QPCs) are a staple of mesoscopic quantum physics, bringing forth the demonstration of noiseless electron flow in ballistic conductors~\cite{Kumar1996}, as well as the observation of fractional electron charges~\cite{Saminadayar1997,dePicciotto1997,Reznikov1999,Dolev2008}, including anyonic statistics~\cite{Bartolomei2020,Nakamura2020} and their scaling dimension~\cite{Veillon2024,Ruelle2024} through noise measurements. Implementing QPCs in graphene has long been an exciting prospect due to its rich quantum Hall (QH) physics, particularly the presence of robust odd- and even-denominator fractional QH states~\cite{Dean2011,Li2017,Zibrov2017}. QPCs have recently been implemented in hexagonal boron nitride (hBN)-encapsulated graphene to realize quantum Fabry-Perot interferometers~\cite{Deprez2021,Ronen2021}, which has led to the demonstration of anyonic statistics in graphene using similar devices in the fractional QH regime~\cite{Werkmeister2024,Samuelson2024}. 
Contrary to GaAs 2DEGs, where the metallic gates of the QPC tend to degrade the electron mobility in the QPC vicinity due to strain~\cite{Davies1994}, the cleanest graphene QPC devices are made with tailored few layered graphite gates~\cite{Cohen2023}, which greatly enhance the mobility~\cite{Zibrov2017}, allowing the precise characterization of the scaling laws predicted by Tomonaga-Lüttinger liquids theory for tunneling into a fractional QH edge channel~\cite{Cohen2023a}. However, at high magnetic field, the combination of low disorder and strong Coulomb interaction can induce electrostatic reconstruction, whereby locally increasing the carrier density at the QPC saddle point becomes energetically favorable~\cite{Cohen2025}. This leads to the formation of a quantum dot which will dictate the transport properties of the QPC. In particular, the shot noise is likely to be affected by the presence of the quantum dot, potentially affecting `collider'-type experiments where quantum statistics of the carriers are obtained from the shot noise~\cite{Kumar1996,Bartolomei2020}. In this letter, we investigate the shot noise of a graphene QPC in the QH regime in presence of reconstruction at the QPC saddle point. We show that the formation of the reconstructed quantum dot leads to a significant increase in the shot noise. We explain this increase by the presence of several edge channels circulating in the quantum dot: Coulomb interactions coupling neighboring channels induce correlations between successive tunneling events across the dot. These memory effects in the charge transfers are characterized by an enhanced shot noise.

\begin{figure*}[ht]
\centering

\includegraphics[width=0.94\textwidth]{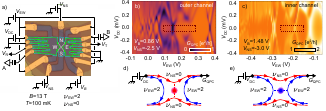}
\caption{\label{fig1} \textbf{(a),} Sample micrograph, indicating the measurement wiring. Dashed white lines: edges of the graphene flake; bright green lines: edge channels paths and chirality. \textbf{(b)} and \textbf{(c),} Conductance across the QPC measured as a function of the bias voltage $\Vdc$ and the east and west gate voltage $V_\mathrm{EW}$, for two settings of the back and north/south gate voltages. The red dashed rectangles indicate the regions corresponding to the data shown in Fig.~\ref{fig2}. \textbf{(d)} and \textbf{(e),} Schematic representations of the edge channels configuration at the QPC corresponding to the data shown in \textbf{(b)} and \textbf{(c)}, respectively. Red arrow: outer edge channel. Blue arrow: inner edge channel. 
} 
\end{figure*}

\begin{figure}[ht]
\centering
\includegraphics[width=0.48\textwidth]{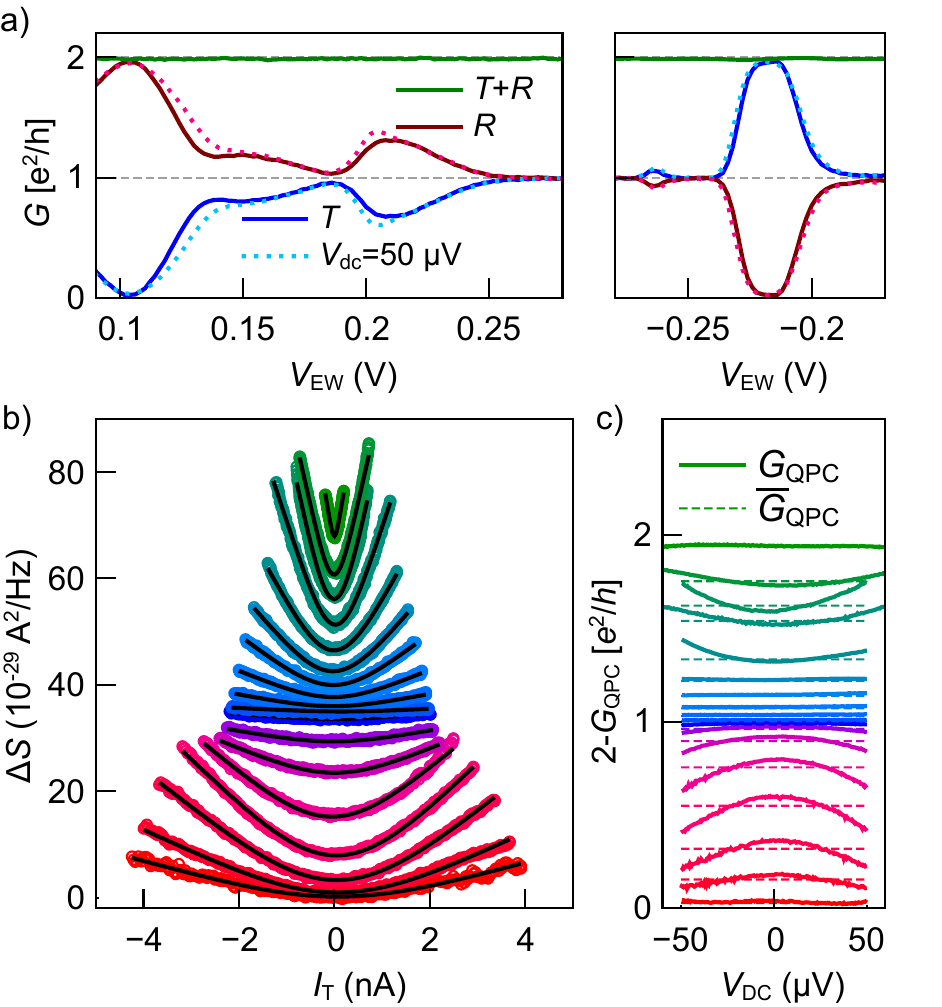}
\caption{\label{fig2} \textbf{(a),} Differential conductances (see text) versus $V_\mathrm{EW}$, in the range framed by the dashed lines in Fig.~\ref{fig1}b) (left panel) and Fig.~\ref{fig1}c) (right panel). Blue (resp. red): transmitted (resp. reflected) conductance. Green: sum of the transmitted and reflected conductances. Full lines: conductances at $V_\mathrm{dc}=0$; dashed lines: conductances at $V_\mathrm{dc}=50~\mu$V. \textbf{(b),} Excess shot noise $\Delta S$ versus transmitted DC current $I_\mathrm{T}$. Symbols: experimental data taken in the regions shown in \textbf{(a)}, for average QPC conductances ranging from $\sim0$ (green) to $\sim2e^2/h$ (red). The curves are shifted vertically for clarity. Black lines: fits using Eq.~\ref{eq:shotnoise} with $F$ as a fitting parameter. \textbf{(c),} reflected QPC conductance versus $\Vdc$, measured simultaneously as the noise data shown in \textbf{(a)}. Dashed lines: average conductance $\bar{G}_\mathrm{QPC}$.
}
\end{figure}

\begin{figure}[ht]
\centering
\includegraphics[width=0.45\textwidth]{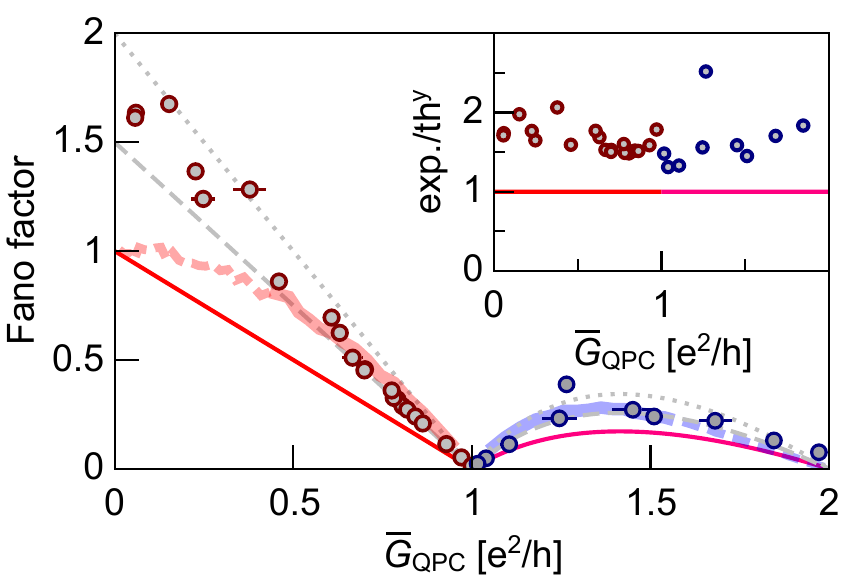}
\caption{\label{fig3} Fano factor versus QPC conductance. Symbols: values extracted from the fits shown in Fig.~\ref{fig2}b (red: outer channel, blue: inner channel). The solid, dashed and dotted lines are the Fano factor predictions with respective prefactors $\alpha=1$, $1.5$ and $2$. The thick lines are the results of the semiclassical model (see text). Inset: ratio between extracted Fano factor and Eq.~\ref{eq:Fano} with $\alpha=1$, as a function of the QPC conductance.
}
\end{figure}

\begin{figure}[ht]
\centering
\includegraphics[width=0.4\textwidth]{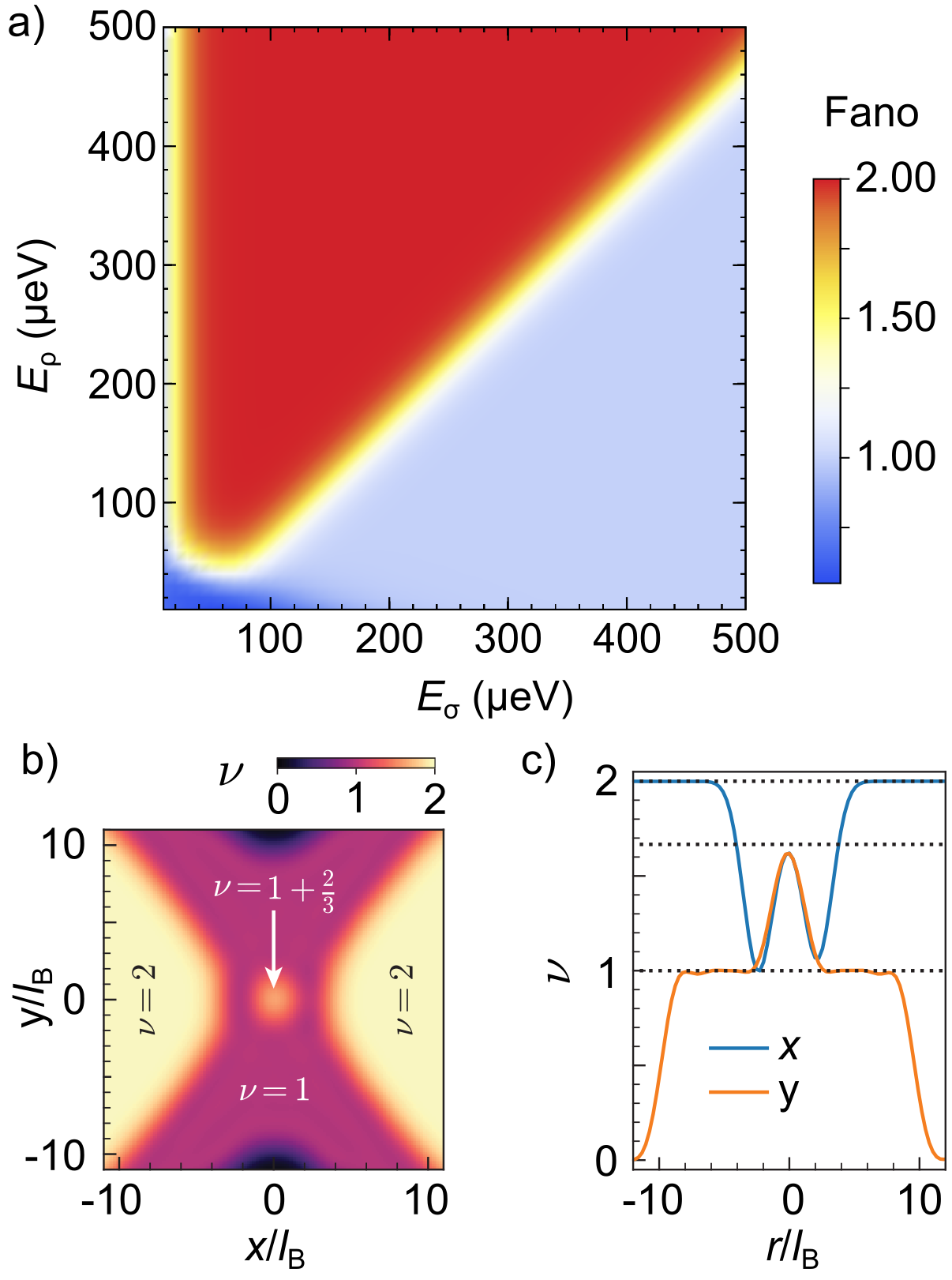}
\caption{\label{fig4} \textbf{(a),} Calculated Fano factor of a $\nu=2/3$ quantum dot versus plasmon modes energy $E_\rho$ and $E_\sigma$ in the plasmon model. \textbf{(b),} Thomas-Fermi calculation of the filling factor variation at the QPC saddle point. $x$ (in units of the magnetic length $l_B=\sqrt{h/2\pi e B}$) corresponds to the W-E direction, $y$ to the S-N direction, and $x=y=0$ to the center of the saddle point. \textbf{(c),} Linecuts of the data of \textbf{(b)} in the $x$ direction at $y=0$ (blue) and in the $y$ direction at $x=0$ (orange). The horizontal dashed lines denote $\nu=1$, $1+2/3$, and $2$.
}
\end{figure}


Fig.~\ref{fig1}a) shows an optical micrograph of the sample, described in details in \cite{Cohen2023}. The QPC is realized by the joint operation of a global graphite back gate (upon which a voltage $V_\mathrm{B}$ is applied) and a graphite top gate divided in four regions: east (E) - west (W), which tune the filling factor $\nu_\mathrm{EW}$ in the bulk, and north (N) - south (S), which tune the filling factor $\nu_\mathrm{NS}$ below the QPC gates. Each pair of gates (E-W, and N-S) is controlled by a single gate voltage: $V_\mathrm{EW}$ and $V_\mathrm{NS}$. At $B=13~$T, $V_\mathrm{B}$ and $V_\mathrm{NS}$ are set to values such that $\nu_\mathrm{NS}=0$, and $V_\mathrm{EW}$ is swept along the $\nu_\mathrm{EW}=2$ plateau, with edge channels flowing clockwise along the edges of the sample. A DC voltage $\Vdc$, along with a small AC excitation $\delta V_\mathrm{AC}$ (not shown in Fig.~\ref{fig1}a) are applied to the upper left contact, and the shot noise is measured on two lines ($A$ and $B$) on both sides of the sample, in auto- and cross-correlations. Transmitted and reflected differential conductances $e^2/h \times dV_\mathrm{T,R}/\delta V_\mathrm{AC}$ are detected on contacts downstream of the noise measurement contacts. The next contacts are connected to a cold ground. The electron density close to the contacts is tuned using the low-resistivity silicon back gate embedded in the substrate, such that there is no reflection at the contacts connected to the cold ground. In all measurements shown here, the dilution refrigerator temperature was fixed and regulated to $100~$mK.

The conductance versus east-west gate voltage and DC voltage maps shown in Fig.~\ref{fig1}b) and c) highlight the two operational settings of the device in which we performed noise measurements, corresponding to a QPC conductance $G_\mathrm{QPC}=e^2/h \times dV_\mathrm{T}/\delta V_\mathrm{AC}$ ranging between $\{0-1\}e^2/h$ (Fig.~\ref{fig1}b), and $\{1-2\}e^2/h$ (Fig.~\ref{fig1}c). Both maps show clear Coulomb diamonds features, as previously reported for these devices~\cite{Cohen2025}, that correspond to the formation of a quantum dot at the QPC saddle point due to electrostatic reconstruction. Remarkably, the resonances in the Coulomb diamonds correspond to a pinch-off of the QPC (both edge channels fully reflected) in Fig.~\ref{fig1}b): $G_\mathrm{QPC}\rightarrow 0$, and to a full opening (both edge channels fully transmitted) in Fig.~\ref{fig1}c): $G_\mathrm{QPC}\rightarrow 2e^2/h$. In both cases, $G_\mathrm{QPC}$ is quantized to $1\times e^2/h$ inside the diamonds, which we interpret as the outer edge channel being fully transmitted, while the inner one is fully reflected. As we show below, our noise measurements confirm this interpretation. From this we infer that the quantum dot only couples to a single channel at a time (outer one in Fig.~\ref{fig1}b), inner one in Fig.~\ref{fig1}c), and that the coupling changes direction between the two channels, as shown in Figs.~\ref{fig1}d) and e). The \textit{outer} channel is coupled in \textit{reflection}: electrons flowing through the quantum dot are reflected at the QPC (Figs.~\ref{fig1}d). The \textit{inner} channel is coupled in \textit{transmission}: electrons flowing through the quantum dot are transmitted through the QPC (Figs.~\ref{fig1}e).

Fig.~\ref{fig2}a) shows the dependence of the transmitted (blue) and reflected (red) differential conductances, in units of $e^2/h$ with $V_\mathrm{EW}$, in the regions delimited with dashed red lines in Fig.~\ref{fig1}b) and c). In the left panel (corresponding to Fig.~\ref{fig1}b), the transmitted conductance varies continuously between $0$ and $1$, while the reflected conductance varies, in a symmetric fashion, between $2$ and $1$. In the right panel, this behavior is inverted, with the transmitted and reflected conductances varying between $1$ and $2$, and $1$ and $0$, respectively. Importantly, in both panels, the two conductances show equal quantization at $1$, over significant ranges of $V_\mathrm{EW}$ corresponding to the Coulomb diamonds in Fig.~\ref{fig1}b) and d). Moreover, their sum (green) is constant and quantized to the Hall conductance $2\times e^2/h$ over the whole range of $V_\mathrm{EW}$. This indicates that charge transport is entirely mediated by the two edge channels, and that the QPC allows fully separating them. Comparing zero bias (full lines) and finite bias ($\Vdc=50~\mu$V, dashed lines) shows that the energy dependence of the QPC transmission is weak in this range. 

The noise measurements performed in these two spans of gate and bias voltages are plotted in Fig.~\ref{fig2}b), as a function of the transmitted DC current $I_\mathrm{T}=\int_0^{\Vdc} G_\mathrm{QPC}(V) dV$. The excess noise $\Delta S$ generated by partitioning at the QPC is extracted from the difference between auto- and cross-correlations \cite{Blanter2000}. The experimental data is shown as symbols. \textcolor{black}{Each curve (shifted vertically for clarity) corresponds to a fixed $V_\mathrm{EW}$, changing the average QPC conductance $\bar{G}_\mathrm{QPC}$ defined as the slope of $I_\mathrm{T}$ versus $\Vdc$ (green: $\approx0$, blue: $1$, red: $\approx2$). The corresponding conductance measurements, performed simultaneously, are shown in Fig.~\ref{fig2}c).} Qualitatively, the noise displays the behavior expected for a QPC: $\Delta S$ increases with $|I_\mathrm{T}|$, with a slope modulated by $\bar{G}_\mathrm{QPC}$. In particular, for $\bar{G}_\mathrm{QPC}$ between 0 and 1 (green to blue), corresponding to the configuration of Fig.~\ref{fig1}b) and d) where the outer channel is partitioned through the QPC, the slope decreases monotonously with $\bar{G}_\mathrm{QPC}$, and vanishes at $\bar{G}_\mathrm{QPC}=1$. For $\bar{G}_\mathrm{QPC}$ between 1 and 2 (blue to red), corresponding to the configuration of Fig.~\ref{fig1}c) and e) where the inner channel is partitioned through the QPC, the slope is maximum at intermediate $\bar{G}_\mathrm{QPC}$, and again decreases as $\bar{G}_\mathrm{QPC}\rightarrow2$. We analyzed quantitatively our data by fitting it with the finite temperature excess shot noise formula~\cite{Blanter2000,SM}:

\begin{equation}
\Delta S = 2 e F \left(  I_\mathrm{T} \times \mathrm{coth}(\frac{e I_\mathrm{T}}{2 \bar{G}_\mathrm{QPC} \kB T}) -2 \bar{G}_\mathrm{QPC} \frac{\kB T}{e} \right),
\label{eq:shotnoise}
\end{equation}

with the Fano factor $F$ as the only fit parameter, $\bar{G}_\mathrm{QPC}$ and $T=100~$mK being fixed. The fits are plotted as black lines in Fig.~\ref{fig2}b) (also shifted vertically for clarity), and show a very good agreement with the measurement data.

The extracted values of $F$ are plotted in Fig.~\ref{fig3} as a function of $\bar{G}_\mathrm{QPC}$. \textcolor{black}{They match well the expected behavior of a QPC sequentially transmitting two channels with respective transmissions $\tau_1=\bar{G}_\mathrm{QPC}$ for $\bar{G}_\mathrm{QPC}\in[0,1]$ and $\tau_2=(\bar{G}_\mathrm{QPC}-1)$ for $\bar{G}_\mathrm{QPC}\in[1,2]$:}

\begin{equation}
F = \alpha \left( (1-\tau_1)+ \frac{\tau_2 (1-\tau_2) }{1+\tau_2}   \right),
\label{eq:Fano}
\end{equation}

albeit with a prefactor $\alpha$ markedly larger than 1: $\alpha\approx 1.5-2$, see inset of Fig.~\ref{fig3}. The typical uncertainty on the extraction of $F$, corresponding to the size of the symbols, is much smaller than the difference between the curve for $\alpha=1$ (full lines) and that for $\alpha=1.5$ (dashed lines). This $1.5-2$ prefactor is observed regardless of the way one calculates and plots the Fano factor for each channel, with very similar results for both channels~\cite{SM}. We observe a similar increase in the noise measured over the full $V_\mathrm{EW}$ and $\Vdc$ range covered by the conductance maps shown in Fig.~\ref{fig1}, see Supplemental Material~\cite{SM}. Importantly, the precise calibration of the noise measurement setup~\cite{SM} and of the injected current, along with the fact that for all $\bar{G}_\mathrm{QPC}$, the thermal rounding of the noise is very well reproduced by a fixed temperature $T=100~$mK equal to that of our fridge, precludes any error larger than $10~\%$ in the extracted value of $\alpha$~\cite{SM}. The measured noise is thus significantly larger than expected, over the whole range of QPC transmission. 
We attribute this enhancement to the presence of the quantum dot at the QPC saddle point. Quantum dot physics can lead to non-Markovian charge transport~\cite{Sukhorukov2001}, where the transfer of one charge across the quantum dot influences that of a subsequent charge. This yields super-Poissonian current fluctuations, that is, shot noise with effective Fano factors larger than 1. This has been extensively studied \cite{Sukhorukov2001, Belzig2005, Thielmann2005a, Onac2006, Weymann2007, Aghassi2008, Carmi2012, Kaasbjerg2015, Zhang2007, Zarchin2007, Fricke2007, Okazaki2013, Harabula2018}, including by some of our coauthors~\cite{Seo2018}; however, most of those mechanisms correspond to high-bias transport, where the quantum dot is either in the sequential tunneling regime, or in the inelastic cotunneling regime. Our data is obtained at energies below the inelastic cotunneling threshold (see Fig.~\ref{fig1}b and c); in this regime, Fano factor enhancements can be observed in quantum dots inside which at least two edge channels circulate~\cite{Choi2015}. The noise enhancement mechanism, recently proposed in ref.~\cite{Frigeri2020}, relies on strong electrostatic interactions between the two edge channels. These lead to two plasmonic eigenstates shared by both edge channels~\cite{Kane1994, Kane1995, Wen1995, Sukhorukov2007, Levkivskyi2008, Ferraro2010, Carrega2012, Bocquillon2013, Rodriguez2020}: a fast, high-energy charge mode, and a slow, low-energy dipolar mode, called neutral mode, or spin mode. If the bias is comparable to the energy of the neutral mode, two competing tunneling processes come into play: i) in the first the tunnelings of two subsequent electrons are independent and leave the state of the quantum dot unaffected, while ii) in the second an electron enters the quantum dot then exits leaving behind an excited neutral plasmonic mode within the dot. If the lifetime of this excitation is long enough, it can then be absorbed by the next tunneling charge, largely enhancing its tunneling probability~\cite{Cavaliere2004,Kim2005,Kim2006}. This induces correlations in the tunneling events, and increases the noise to effective Fano factors up to 2~\cite{Frigeri2020}. \textcolor{black}{Note that while this effect is more commonly observed through the doubling~\cite{Choi2015} or tripling~\cite{Yang2024} of the Aharonov-Bohm frequency in quantum Hall Fabry-Perot interferometers including at least two edge channels, it also appears in the shot noise~\cite{Choi2015,Frigeri2020}, and should be present even in the case of large charging effects.}
Fig.~\ref{fig3} shows a comparison of our results with a semiclassical realization of this process that ignores the microscopics of the quantum dot, in particular its edge channel structure. In this model (thick lines in Fig.~\ref{fig3}), the tunneling correlations are simplified to the extreme case: when a charge goes through the dot with a probability $p_\mathrm{t}$, the next one goes through the dot with a probability $1$. After this second charge tunneling, the process is reset. The evolution of the Fano factor with the probability $p_\mathrm{t}$ can then be understood in the following way. For low $p_\mathrm{t}$ (corresponding to large $\tau_1$ or low $\tau_2$), the train of charges arriving on the dot generates individual kinks of transmitted charge $2e$, leading to a Fano factor close to $2$. Inversely, for $p_\mathrm{t}$ close to $1$ ($\tau_1 \rightarrow 0$ or $\tau_2 \rightarrow 1$), our model leads to individual kinks of charge $e$ in the backscattered current and hence a Fano factor close to $1$. The agreement between the experimental data and this model, without any fit parameters, is very reasonable in the range $0.5e^2/h<G_\mathrm{QPC}<1.5e^2/h$ (full lines), corresponding to $p_\mathrm{t}<0.5$. For larger $p_\mathrm{t}$, sizeable deviations appear (dashed lines), and the calculated Fano factor reaches $1$ for $G_\mathrm{QPC}\rightarrow0$. This is due to the fact that in our experiment, unity transmission across the dot correspond to resonant tunneling rather than transparent barriers. Note that this model relies on a large number of independent computations for each value of $p_\mathrm{t}$; the fluctuations shown in Fig.~\ref{fig3} correspond to a finite number of such computations.

According to the above discussion, the quantum dot requires two circulating edge channels for charge tunneling correlations to appear. Our conductance measurements show that the local filling factor at the QPC saddle point is at least $1$ (purple region in Fig.~\ref{fig1}d and e), and increases to give rise to the quantum dot. In order to accommodate two integer edge channels, the filling factor in the dot should reach $\nu=3$, which is unlikely because of the very large cyclotron gap between the zeroth and first Landau levels of graphene. Thus, the filling factor in the dot is likely non-integer, potentially giving rise to fractional quantum Hall edge channels circulating in the dot. A good candidate is a local $\nu=1+2/3$ state, which, in absence of disorder, hosts two counterpropagating edge states on top of an inert background at $\nu=1$: a downstream integer one, and an upstream fractional one with conductance $1/3\times e^2/h$~\cite{MacDonald1990}. We have adapted the plasmon model of ref.~\cite{Frigeri2020} to the $\nu=2/3$ case of counterpropagating channels with conductances $1\times e^2/h$ and $-1/3\times e^2/h$ (see Supplemental Material~\cite{SM}) to calculate the Fano factor through the quantum dot in the regime of small dot transmissions. \textcolor{black}{The two main parameters of the model are the energies of the charge and dipole plasmon modes, respectively $E_\rho$ and $E_\sigma$, with $E_\rho$ being typically larger than the dot charging energy ($\approx300~\mu$eV from the conductance measurements shown in Fig.~\ref{fig1}, see also End Matter), and $E_\sigma\lesssim0.5E_\rho$~\cite{Rodriguez2020}. Setting $T=100~$mK and $\Vdc=50~\mu$V, we plot in Fig.~\ref{fig4} the evolution of the Fano factor versus $E_\rho$ and $E_\sigma$.} As soon as $E_\sigma>\kB T$, irrespective of the value of $E_\rho$, the Fano factor exceeds $1$ and increases with $E_\sigma$ until it saturates to $2$, qualitatively matching our observations. We have validated our hypothesis of the local $\nu=1+2/3$ state by performing Thomas-Fermi calculations of the filling factor at the QPC saddle point~\cite{SM}, (see Fig.~\ref{fig4}b and c). In a previous work realized on the same device~\cite{Cohen2025}, this method had already shown that it was possible to generate a fractional island at the QPC saddle point. With electrostatic parameters (gate voltages, and hBN thickness) matching that of our experiment, we show that the filling factor at the QPC indeed increases from a local $\nu=1$ background to a small zone at $\nu=1+2/3$ at the center of the saddle point, giving rise to an antidot-like structure~\cite{Sim2003}. Importantly, for the noise enhancement to occur in both outer and inner channels of bulk $\nu=2$, the two channels circulating in the quantum dot require spin polarizations matching that of each edge channels of $\nu=2$. The $\nu=2/3$ state in graphene has be shown to be fully spin polarized at large magnetic fields~\cite{Zeng2019}. Near the QPC, the $\nu=2$ edge is defined by the $\nu=0$ region below the N and S gates, leading to a canting of the spin polarizations of both edge channels with respect to the magnetic field~\cite{Takei2016}. This canting can thus reduce the energy cost for charges from either edge channel to tunnel into the spin-polarized quantum dot.

In conclusion, we have shown that self-reconstructed quantum dots at the saddle point of graphene QPCs can lead to a substantial increase in the shot noise. This increase, which we attribute to correlated charge tunneling across the dot mediated by electrostatically coupled counter-propagating QH edge channels circulating in the dot, can be detrimental to future experiments based on shot noise generated at a graphene QPC to probe the unique properties of edge excitations in the QH regime.

\begin{acknowledgments}
 This work was funded by the ERC (ERC-2018-STG \textit{QUAHQ}), by the “Investissements d’Avenir” LabEx PALM (ANR-10-LABX-0039-PALM), and by the Region Ile de France through the DIM QUANTIP. O.M. acknowledges funding from the ANR (ANR-23-CE47-0002 CRAQUANT). J.F. acknowledges support from the UC Berkeley College of Engineering Jane Lewis Fellowship. K.W. and T.T. acknowledge support from the JSPS KAKENHI (Grant Numbers 21H05233 and 23H02052) , the CREST (JPMJCR24A5), JST and World Premier International Research Center Initiative (WPI), MEXT, Japan. M.S acknowledges the support from the project PRIN2022 2022-PH852L(PE3) TopoFlags- “Non reciprocal supercurrent and topological transition in hybrid Nb-InSb nanoflags” funded by the European community Next Generation EU within the programme “PNRR Missione Componente Investimento Fondo per il Programma Nazionale di Ricerca e Progetti di Rilevante Interesse Nazionale (PRIN)”. The authors warmly thank C. Altimiras and A. Assouline for enlightening discussions. 
\end{acknowledgments}

\section{End Matter}

\begin{figure}[ht]
\centering
\includegraphics[width=0.35\textwidth]{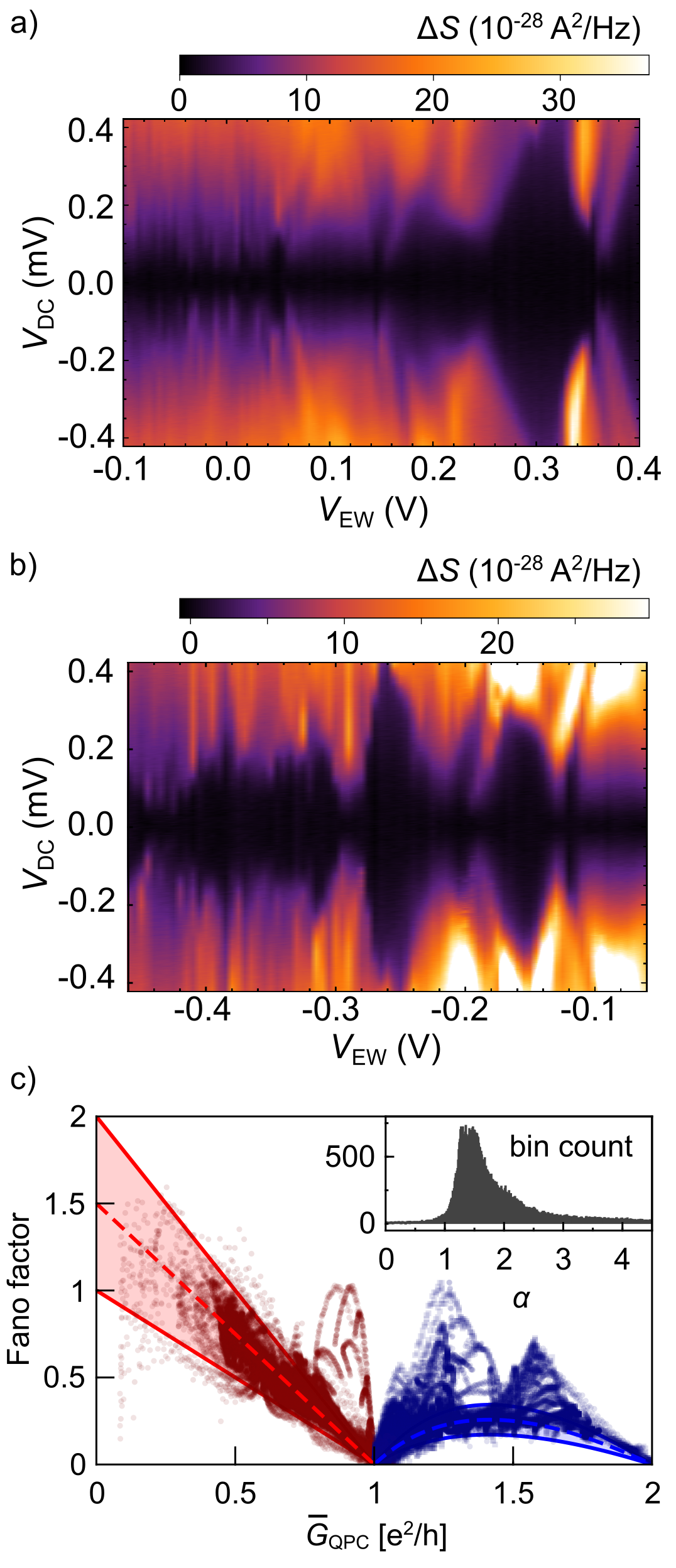}
\caption{\label{figEM1} \textbf{(a)} and \textbf{(b),} Measured excess noise versus $\Vdc$ and $V_\mathrm{EW}$ in the range corresponding to the conductance plots shown in Fig.~\ref{fig1}. \textbf{(c),} Fano factor versus QPC conductance for the two noise datasets: red symbols for \textbf{(a)}, dark blue symbols for \textbf{(b)}. The red (resp. blue) shaded areas correspond to the Fano factor predictions with $\alpha$ between $1$ and $2$ (dashed line: $\alpha=1.5$). Inset: bin count of the values of $\alpha$ extracted from the data in the main panel (bin size: $0.1$).
}
\end{figure}

\subsection{Experimental setup}

The geometry described in Fig.~\ref{fig1} relies on the chirality of the \QH effect to physically separate the different measurement terminals. In particular, the use of cold grounds on both sides of the samples allows isolating the QPC from the back-action of the noise measurement lines, such that their noise (thermal noise and amplifier current noise) does not loop back to the QPC~\cite{SM}, and are considered a constant offset in the noise. The noise measurement lines consist of a shunting RLC tank (with resonance frequency at $2.8~$MHz) followed by an amplification line made of low-noise cryogenic amplifiers and room-temperature amplifier. The setup and calibration procedure for the RLC tank parameters and the amplifiers gains are detailed in the Supplemental Material~\cite{SM}. The calibration procedure, based on thermal noise measured at different mixing chamber temperatures between $20~$mK and $220~$mK, was repeated three times over the two months during which the device was measured. The calibration data was independently analyzed by different authors, yielding similar numbers. The overall uncertainty on the gains is smaller than $6~\%$~\cite{SM}. The fridge temperature was regulated to $100~$mK so as to fix the electron temperature in the noise versus current fits. Additional measurements performed at lower fridge temperature (20~mK) yielded electron temperatures of $\sim60~$mK with similar Fano factor enhancements~\cite{SM}. The conductance measurements were done by applying ac currents at different frequencies (all below 10~Hz) on the various input/outputs of the sample. The current excitations are converted into ac voltages through the quantized hall resistance of the sample, with typical amplitudes of $4-5~\mu$V.

\subsection{Noise measurements at high bias}

The enhanced noise prefactor is also observed in the noise measured over the full $V_\mathrm{EW}$ and $\Vdc$ range, covered by the conductance maps shown in Fig.~\ref{fig1}. Fig.~\ref{figEM1} shows the corresponding excess noise data, with markedly large amplitudes (about an order of magnitude larger than the data shown in Fig.~\ref{fig2}) in the regions outside of the Coulomb diamonds appearing in Fig.~\ref{fig1}. We compute the Fano factor by dividing the measured noise by the expected Poisson noise $2e |I_\mathrm{T}|$. The contribution of the thermal rounding is taken into account by adding to $\Delta S$ a correction factor $4e^2/h\times\tau_{\{1,2\} }(1-\tau_{\{1,2\} }) \kB T$. The obtained values of the Fano factor are plotted in Fig.~\ref{figEM1}c) as a function of $\bar{G}_\mathrm{QPC}$, for all points appearing in the maps except for those at $|\Vdc|<20~\mu$V. The data generally follows the Fano factor formula of main text Eq.~2, with clear shot noise suppression for integer $\bar{G}_\mathrm{QPC}$. However, the corresponding values of $\alpha$ are clearly larger than 1, with a majority of points lying in between $\alpha=1$ and $2$ (red and blue shaded areas). The inset shows the distribution of the extracted $\alpha$ over all measured data (approx. $72000$ points), with a very pronounced peak at $\alpha\approx 1.5$. The distribution is clearly skewed towards large values of $\alpha$, with a very small minority of data corresponding to $\alpha=1$. The analysis of the linecuts at fixed $V_\mathrm{EW}$, shown in the Supplemental Material~\cite{SM}, indicates that the measured noise is systematically larger than the one expected from a scattering formalism, even including the energy dependence of the conductance.

\subsection{Consistency of the models}

In the plasmon model, the charge and dipole mode energies $E_\rho$ and $E_\sigma$ are related to the dot charging energy $E_c^\ast$ through the relation $E_c^\ast=\frac{3}{4}E_\rho+\frac{1}{4}E_\sigma$~\cite{SM}. The conductance measurements shown in Fig.~\ref{fig1} provide an estimate $E_c^\ast\approx300~\mu$eV, thus yielding $E_\rho \approx350-400~\mu$eV, and $E_\sigma \approx100-200~\mu$eV. $E_\rho$ and $E_\sigma$ are related to the dot's perimeter $L$ and the plasmon edge velocities $v_{\rho,\sigma}$ through the relation $E_{\rho,\sigma}=\hbar v _{\rho,\sigma}/L$. $L$ can be estimated from the Thomas-Fermi calculation to $L\approx2\pi \times 2l_B$, yielding $v_\rho\approx48~$km$/$s. This value is about 1.5 times smaller than the typical Fermi velocities obtained in recent quantum Hall Fabry-Perot interferometer experiments~\cite{Deprez2021,Ronen2021}, and about 2 times smaller than the charge plasmon velocities obtained in GaAs experiments~\cite{Rodriguez2020}. Similarly, the island self-capacitance estimated from the conductance measurements can be compared with the Thomas-Fermi calculations. The self-capacitance of about $0.3~$fF obtained from the $E_c^\ast\approx300~\mu$eV charging energy is much larger than a naïve disk plate capacitor model with a radius of $2l_B$. This discrepancy can be qualitatively explained by the unusually soft confining potential at the QPC that gives rise to the island. Because the potential is shallow, the cheapest way to add charge is to let the island expand rather than increase its local density, which boosts the effective self-capacitance far beyond the metallic-disk value and correspondingly suppresses the charging energy. Because the edge term dominates this charging energy, the same soft confinement also lowers the plasmon velocity, possibly accounting for the smaller velocity compared with earlier studies on sharply defined macroscopic edges. In addition, the conductance data show that the electrostatic lever arm of the dc voltage is unusually large, $\alpha\approx300$, denoting an important contribution from the neighboring edge channels to the overall island self capacitance.

\end{document}


\title{Supplemental Material for ''Enhanced shot noise in graphene quantum point contacts with electrostatic reconstruction''}

\author{M. Garg}\thanks{Equal contribution.}
\affiliation{Universit\'e Paris-Saclay, CEA, CNRS, SPEC, 91191 Gif-sur-Yvette cedex, France
}
\affiliation{Department of Physics, Indian Institute of Technology Roorkee, Uttarakhand 247667, India
}

\author{O. Maillet}\thanks{Equal contribution. Contact: olivier.maillet@cea.fr}
\affiliation{Universit\'e Paris-Saclay, CEA, CNRS, SPEC, 91191 Gif-sur-Yvette cedex, France
}

\author{N. L. Samuelson}
\affiliation{Department of Physics, University of California at Santa Barbara, Santa Barbara CA 93106, USA}

\author{T. Wang}
\affiliation{Department of Physics, University of California at Berkeley, Berkeley, CA 94720, USA}
\affiliation{Material Science Division, Lawrence Berkeley National Laboratory, Berkeley, CA 94720,
USA}

\author{J. Feng}
\affiliation{Graduate Group in Applied Science \& Technology, University of California, Berkeley, California 94720, USA}

\author{L. A. Cohen}
\affiliation{Department of Physics, University of California at Santa Barbara, Santa Barbara CA 93106, USA}

\author{A. Zhang}
\affiliation{Universit\'e Paris-Saclay, CEA, CNRS, SPEC, 91191 Gif-sur-Yvette cedex, France
}

\author{K. Watanabe}
\affiliation{Research Center for Electronic and Optical Materials, National Institute for Materials Science, 1-1 Namiki, Tsukuba 305-0044, Japan
}
\author{T. Taniguchi}
\affiliation{Research Center for Materials Nanoarchitectonics, National Institute for Materials Science,  1-1 Namiki, Tsukuba 305-0044, Japan
}

\author{P. Roulleau}\thanks{Contact: preden.roulleau@cea.fr}
\affiliation{Universit\'e Paris-Saclay, CEA, CNRS, SPEC, 91191 Gif-sur-Yvette cedex, France
}

\author{M. Sassetti}
\affiliation{
Dipartimento di Fisica, Universit\`a di Genova, Via Dodecaneso 33, 16146, Genova, Italy
}
\affiliation{
SPIN-CNR, Via Dodecaneso 33, 16146, Genova, Italy
}

\author{M. Zaletel}
\affiliation{Department of Physics, University of California at Berkeley, Berkeley, CA 94720, USA}
\affiliation{Material Science Division, Lawrence Berkeley National Laboratory, Berkeley, CA 94720,
USA}

\author{A. F. Young}
\affiliation{Department of Physics, University of California at Santa Barbara, Santa Barbara CA 93106, USA}

\author{D. Ferraro}
\affiliation{
Dipartimento di Fisica, Universit\`a di Genova, Via Dodecaneso 33, 16146, Genova, Italy
}
\affiliation{
SPIN-CNR, Via Dodecaneso 33, 16146, Genova, Italy
}

\author{P. Roche}
\affiliation{Universit\'e Paris-Saclay, CEA, CNRS, SPEC, 91191 Gif-sur-Yvette cedex, France
}

\author{F.D. Parmentier}\thanks{Contact: francois.parmentier@phys.ens.fr}
\affiliation{Universit\'e Paris-Saclay, CEA, CNRS, SPEC, 91191 Gif-sur-Yvette cedex, France
}
\affiliation{Laboratoire de Physique de l’Ecole normale sup\'erieure, ENS, Universit\'e PSL,
CNRS, Sorbonne Universit\'e, Universit\'e Paris Cit\'e, F-75005 Paris, France
}

\date{\today}

\maketitle

\section{Sample fabrication}
The graphite top gate was patterned via AFM anodic oxidation lithography (as outlined in ref.~\cite{Cohen2023}, a prior study of the same device, which gives more detailed fabrication process parameters). Then, the van der Waals heterostructure was stacked via a dry transfer process using a polycarbonate film on a dome shaped PDMS to pick up each successive layer. The stack was deposited onto a doped silicon substrate with a 285nm-thick thermally-grown oxide layer. 30kV e-beam lithography was then used to define the pattern of an e-beam evaporated 40nm aluminum hard mask, and the device outline was etched via an inductively-coupled CHF3/O2 plasma etch. The aluminum mask was removed with a dilute solution of TMAH, and a PMMA mask written via a second EBL step before e-beam evaporation of Cr/Pd/Au (3/15/185 nm) edge contacts. Finally, another PMMA mask was used to define narrow trenches which were etched into each contact via an additional CHF3/O2 plasma etch step.

\section{Measurement setup}

\subsection{Conductance measurements}
\begin{figure*}[h!]
\centering
\includegraphics[width=0.75\textwidth]{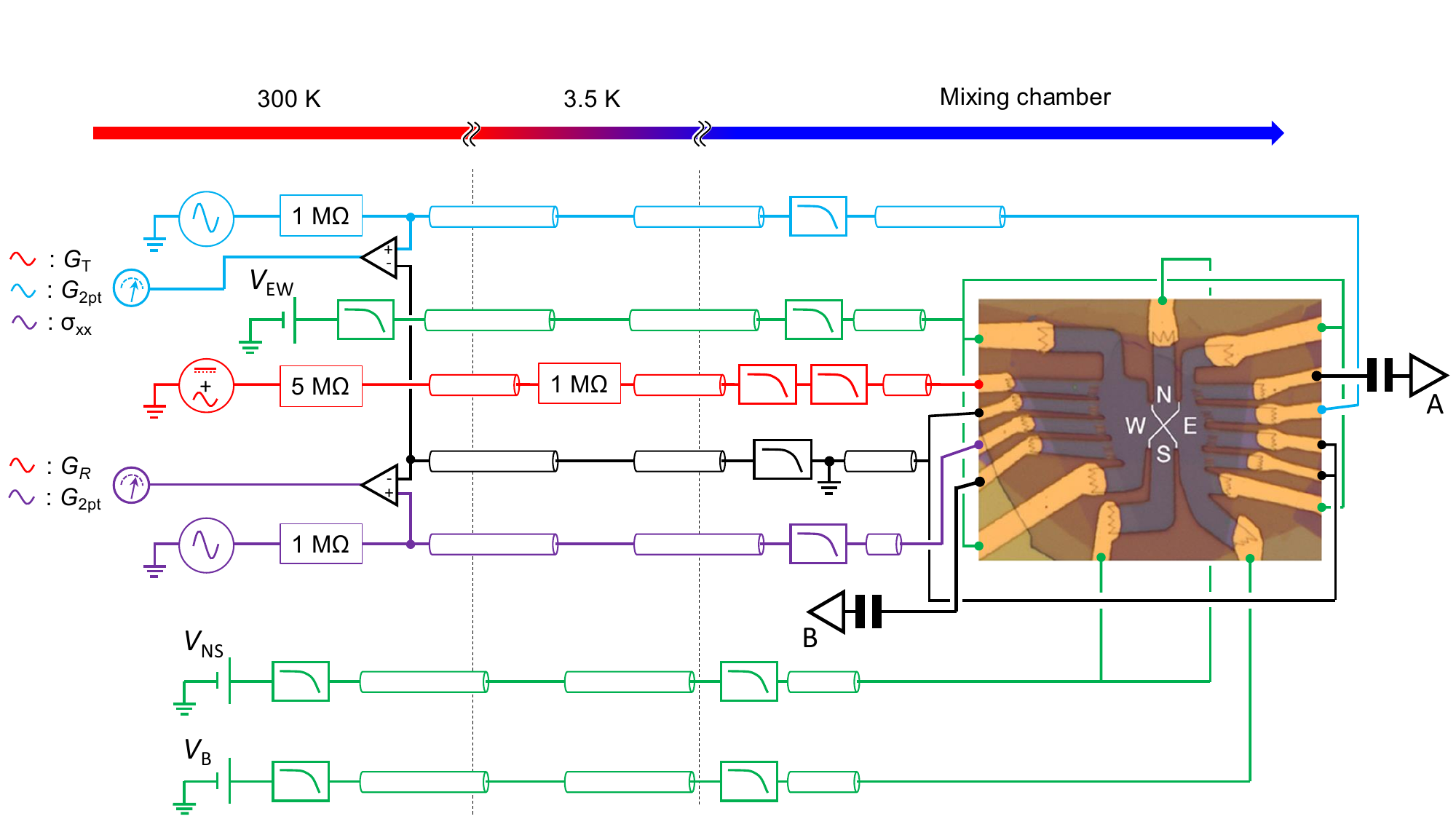}
\caption{\label{figsup-conductances} Layout of the wiring for the conductance measurements. Lines are color-coded (blue: T-side conductance; green: gates; red: dc current feed; purple: R-side conductance; black: cold ground).}
\end{figure*}
A detailed description of the conductance measurements is shown in Supplemental Fig.~\ref{figsup-conductances}. The measurements were performed using lock-in techniques at low frequency, below 10~Hz. The sample is current biased (red line) through a 6 M$\Omega$ resistor (which  includes a $1~$M$\Omega$ series bias resistor thermally anchored to the $3.5~$K stage of our dilution refrigerator); this bias is converted into a perfect voltage bias $\Vdc$ through the sample's Hall resistance $h/\nu e^2$. All lines, including current feed and gates  are heavily filtered at the mixing chamber stage of our dilution refigerator using cascaded \textit{RC} filters. The effect of those filters (both in terms of series resistance and capacitive cutoff) are taken into account in our data. All measurements are performed using differential amplifiers (CELIANS EPC-1B) referenced to the cold ground (black in Fig.~\ref{figsup-conductances}) The latter is directly connected (both electrically and thermally) to the mixing chamber stage. The cold ground is used to isolate the sample from any back-action of the noise measurement chain. In particular, the current noise of the amplifiers flowing back to the sample flow into a constant impedance given by that of the RLC tank (see below) in parallel with the resistance between the noise measurement contact and the downstream cold ground, equal to the transverse resistance $h/\nu e^2$. This resistance is thus independent of the QPC transmission, and thus on the dc and gate voltages provided $\nu_\mathrm{EW}$ remains constant.

\subsection{Noise measurements: setup and calibration}

\begin{figure}[h!]
    \centering
    \includegraphics[width=0.7\textwidth]{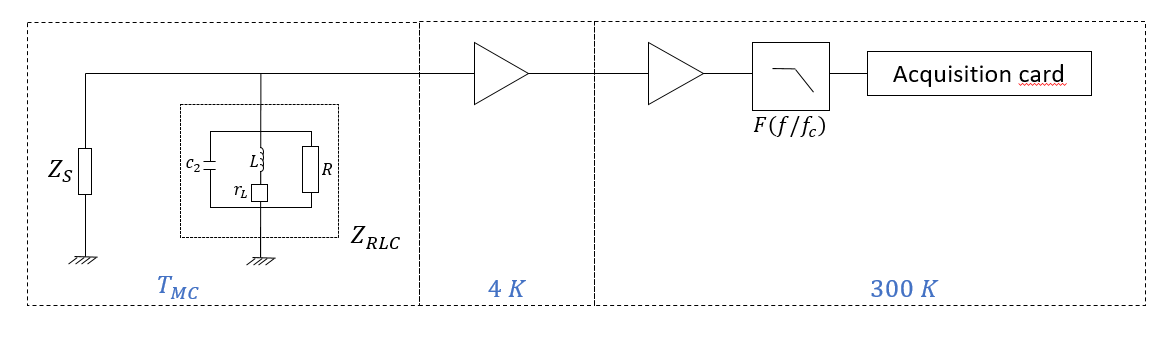}
    \caption{Noise measurement circuit from the sample to the acquisition card. The resonator is made with a capacitor $c_2$, an inductor $L$ with an effective resistance $r_L$ and a resistance $R$ which represents the losses on the circuit. }
    \label{fig:calib_circuit}
\end{figure}

\begin{figure}[h!]
    \centering
    \includegraphics[width=0.8\textwidth]{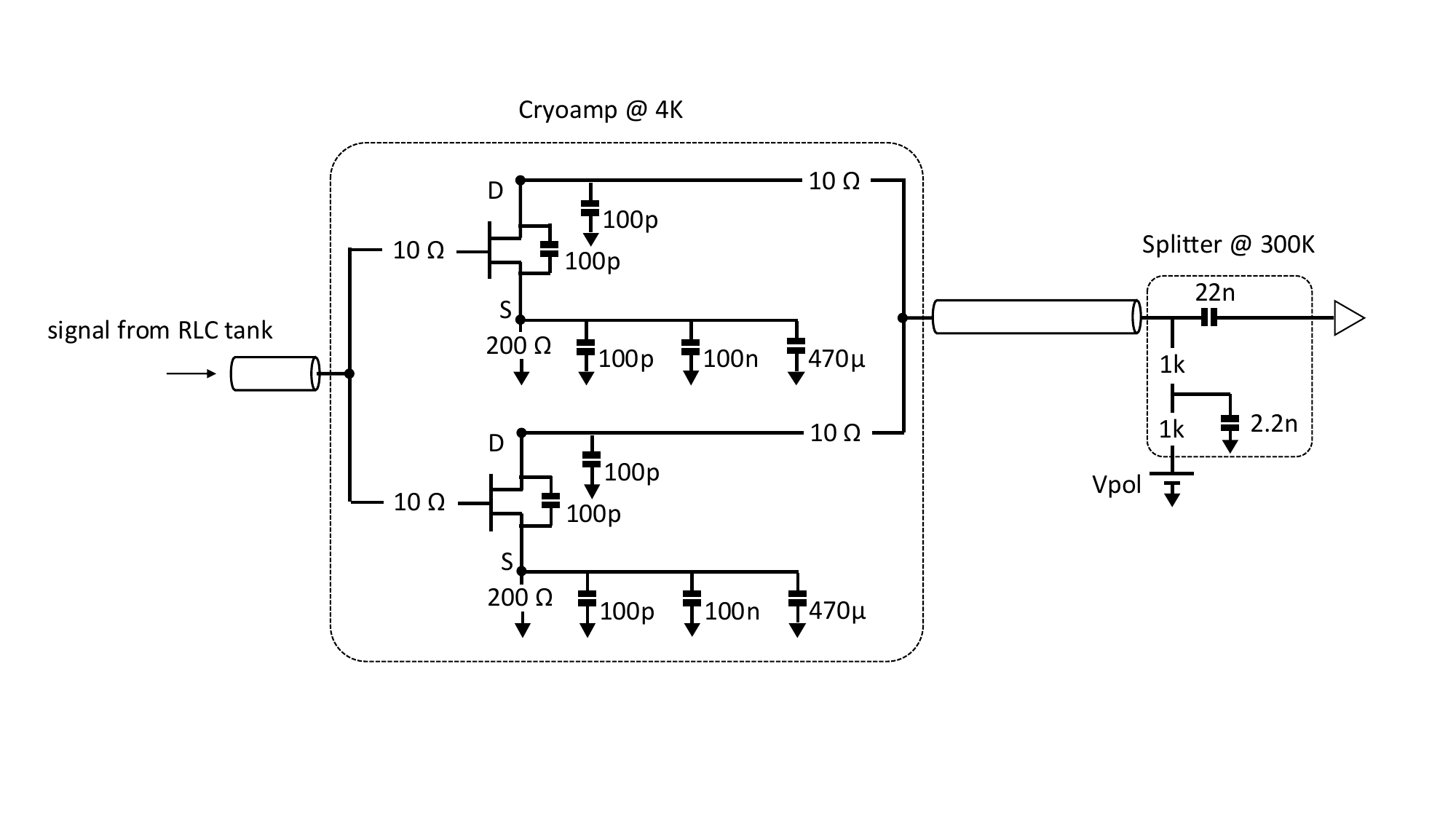}
    \caption{Circuit schematics of the cryoamps, based on two parallel ATF 34143 HEMTs.}
    \label{fig:cryoamps}
\end{figure}

\begin{figure*}[h!]
    \centering
    \includegraphics[width=0.97\textwidth]{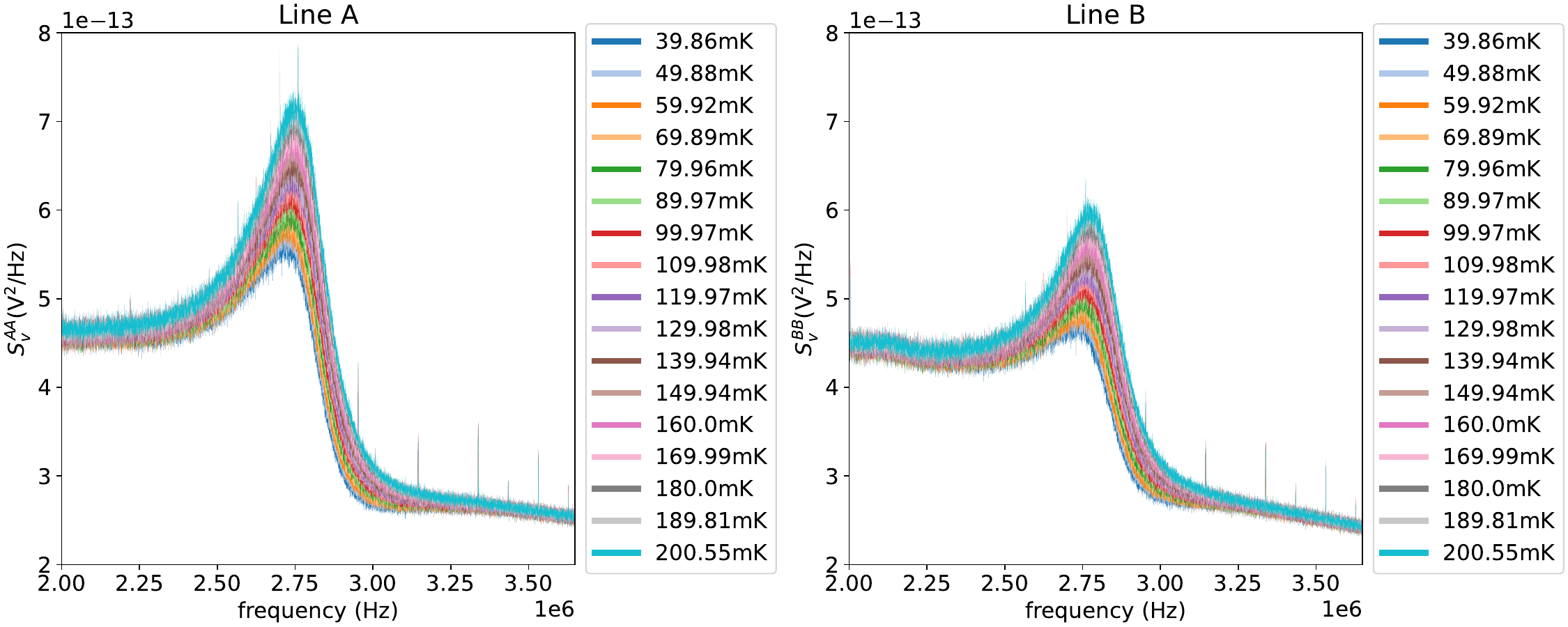}
    \includegraphics[width=0.97\textwidth]{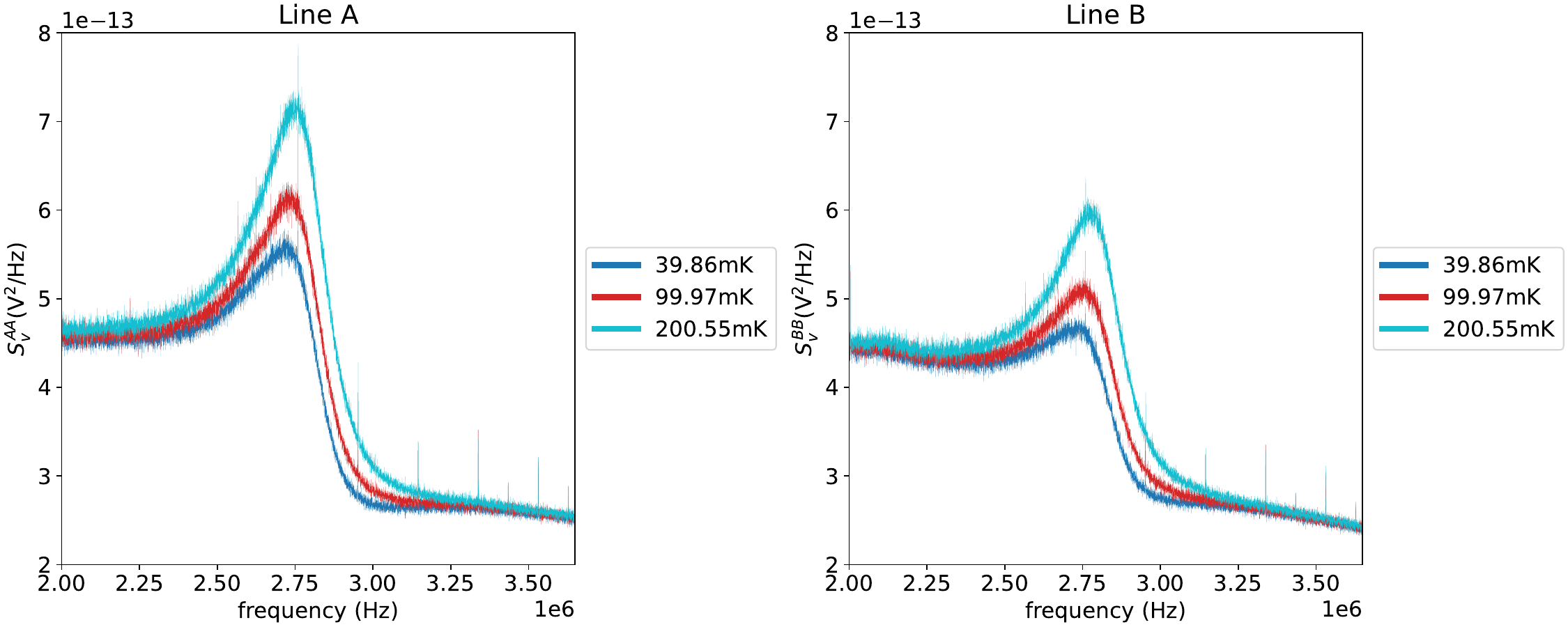}
    \caption{Raw noise spectra for line A (left) and B (right) for different temperatures ranging from 39 mK to 200 mK. Data taken at $\nu_\mathrm{EW}=2$, at $B=13~$T. Three temperatures are selected and plotted below for the sake of visibility. }
    \label{fig:calib_raw}
\end{figure*}

The noise measurement circuit consists of two independent filtering and amplification lines (A and B), schematized in Fig.~\ref{fig:calib_circuit}. In each line, a RLC tank is connected in parallel to the sample and anchored to the mixing chamber stage of our refrigerator. The tank transforms the current fluctuations of the sample into voltage fluctuations that are amplified by a double stage consisting in home-made cryoamps followed by room-temperature follower amplifiers (NF-SA-220 F5). Fig.~\ref{fig:cryoamps} shows the circuit schematics of our cryoamps, based on two parallel commercial ATF 34143 HEMTs. The biasing scheme is similar to the one described in \cite{DiCarlo2006}. The cryoamp voltage gain is typically $8-9$, and that of the room-temperature follower amplifier is 400. Typical voltage noises are $0.28~$nV$/\sqrt{\mathrm{Hz}}$. After amplification, the voltage in each line recorded by a dual input Spectrum GMBH digitizer board with which we compute the auto and cross spectra ${S}_v^{AA}(f),~{S}_v^{BB}(f),~{S}_v^{AB}(f)$ that are then averaged over a frequency bandwidth $\Delta f=[2.2-3.2]$~MHz encompassing the RLC tank resonance. To extract the current noise of the sample $S_{i,sample}$ from the averaged auto and cross spectra $\bar{S}_v^{AA},~\bar{S}_v^{BB},~\bar{S}_v^{AB}$, one need to accurately describe and determine all the parameters of the amplification chain. We start from the following expressions:


\begin{align}
        &\bar{S}_v^{AA}=\frac{1}{\Delta f}\int_{\Delta f}{S}_v^{AA}(f)=G_A^2 \times \frac{1}{\Delta f}\int_{\Delta f} df F( \frac{f}{f_{c}^A} ) \left[ S_{v,amp}^A +
    \left\lvert Z_{//}^A\right\rvert^2 \left( S^A_{i,amp} +4k_BTRe\left(\frac{1}{Z_{RLC}^A}\right)+ S_{i,sample} \right) \right]\\
        &\bar{S}_v^{BB}=\frac{1}{\Delta f}\int_{\Delta f}{S}_v^{BB}(f)=G_B^2 \times \frac{1}{\Delta f}\int_{\Delta f} df F( \frac{f}{f_{c}^B} ) \left[ S_{v,amp}^B +
    \left\lvert Z_{//}^B\right\rvert^2 \left( S^B_{i,amp} +4k_BTRe\left(\frac{1}{Z_{RLC}^B}\right)+ S_{i,sample} \right) \right]\\
    &\bar{S}_v^{AB}=\frac{1}{\Delta f}\int_{\Delta f}{S}_v^{AB}(f)=-G_A G_B\times \frac{1}{\Delta f}\int_{\Delta f} df \sqrt{F( \frac{f}{f_{c}^A} )F( \frac{f}{f_{c}^B} )} \left[ Z_{//}^A\left( Z_{//}^B \right)^\ast S_{i,sample} \right]
    \label{eq:thnoise_meas}
\end{align}
with $G_{A/B} = G_{A/B}^{RT} \times G_{A/B}^{HEMT}$ the overall gain of the amplification chain from room temperature amplifiers (of gain $G_{A/B}^{RT}$) and cryoamps (of gain $G_{A/B}^{HEMT}$) at $4 \, \mathrm{K}$, $F(f/f_{c}^{A/B})$ a second order low-pass filter function corresponding to the antialiasing filters (see Supplemental Fig.~\ref{fig:calib_circuit}), $S^{A/B}_{i,amp}$ and $S^{A/B}_{v,amp}$ the current and voltage noises of the amplifier A or B, $4k_B T Re(\frac{1}{Z_{RLC}^{A/B}})$ the thermal noise of the LRC circuit resonator, $Z_{//}^{A/B}$ the parallel impedance of the LRC circuit resonator and the sample, $\Delta f$ the integration bandwidth, and  $S_{i,sample}$ the current noise of the sample, which includes its thermal noise written as $4k_B\Delta T \times \nu \times e^2/h$ at filling factor $\nu$ under quantum Hall regime. To determine the parameters of the RLC circuit resonator as well as the voltage and current of our amplifiers, we use a temperature calibration method. We measure the equilibrium noise from $10$ to $200\,\mathrm{mK}$ at different filling factors $\nu$. Supplemental Fig.~\ref{fig:calib_raw} shows typical raw spectra obtained from this calibration. We show only the temperature ranging from $40$ to $200\, \mathrm{mK}$, as under $40\, \mathrm{mK}$ electrons are not well thermalized.  

According to Supplemental Eq.~\ref{eq:thnoise_meas}, the noise includes a temperature independent part due to the amplifier's noise. In order to first extract the values of the gain and the RLC circuit parameters, we remove to each spectra shown in Supplemental Fig.~\ref{fig:calib_raw} a reference spectrum made of the average of all spectra for temperatures ranging from $50~$mK to $200~$mK:

\begin{equation}
    \Delta S_v^{AA/BB}(f,T) = S_v^{AA/BB}(f,T)-\frac{1}{N_T}\sum_{k=1}^{N_T}S_v^{AA/BB}(f,T_k)
    \label{eq:deltaSv}
\end{equation}
The obtained spectra, shown in Supplemental Fig.~\ref{eq:calib_thermalnoise}, are then fitted by the expression below:


\begin{equation}
    \Delta {S}_v^{AA/BB}(f)=G_{A/B}^2\ F \left(\frac{f}{f_c^{A/B}}\right) 4k_B\Delta T \left\lvert Z_{//}^{A/B}(f) \right\rvert^2 \left[ Re \left( \frac{1}{Z_{RLC}^{A/B}(f)} \right) + \nu \times e^2/h  \right]
    \label{eq:calib_thermalnoise}
\end{equation}

Fitting Eq. \ref{eq:calib_thermalnoise} from the measured noise gives us the parameter $R_{A/B}$, $c_{2_{A/B}}$, $L_{A/B}$, $r_{L_{A/B}}$ shown in Supplemental Fig. \ref{fig:calib_circuit} and the gain $G_{A/B}$. The fitting result is shown in Supplemental Fig. \ref{fig:calib_diff}. This calibration can then be applied to the measurements, by simply dividing the integrated noise measured at any given gate or bias voltage by a fixed conversion factor which only depends on the bulk filling factor of the sample:

\begin{align}
    S_{i,sample} &= \frac{ \bar{S}_v^{AA/BB}}{\frac{G_{A/B}^2}{\Delta f} \int_{\Delta f}df F(\frac{f}{f_c^{A/B}}) \left\lvert Z_{//}^{A/B}(f,\nu)\right\rvert^2 }\\
    &=\frac{ \bar{S}_v^{AB}}{-\frac{G_A G_B}{\Delta f}\int_{\Delta f} df \sqrt{F( \frac{f}{f_{c}^A} )F( \frac{f}{f_{c}^B} )} \left[ Z_{//}^A(f,\nu)\left( Z_{//}^B(f,\nu) \right)^\ast  \right]}
    \label{eq:calib_thermalnoise2}
\end{align}

\begin{figure*}[h!]
    \centering
    \includegraphics[width=0.97\textwidth]{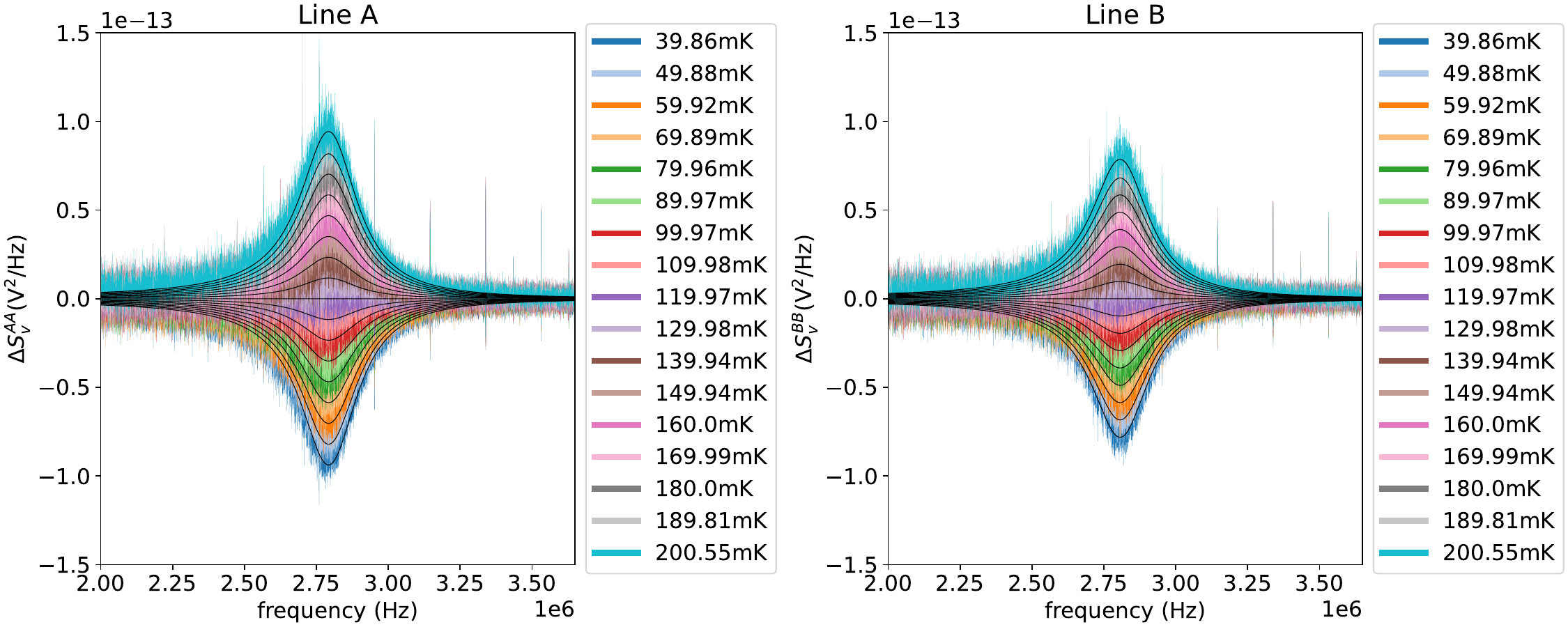}
    \includegraphics[width=0.97\textwidth]{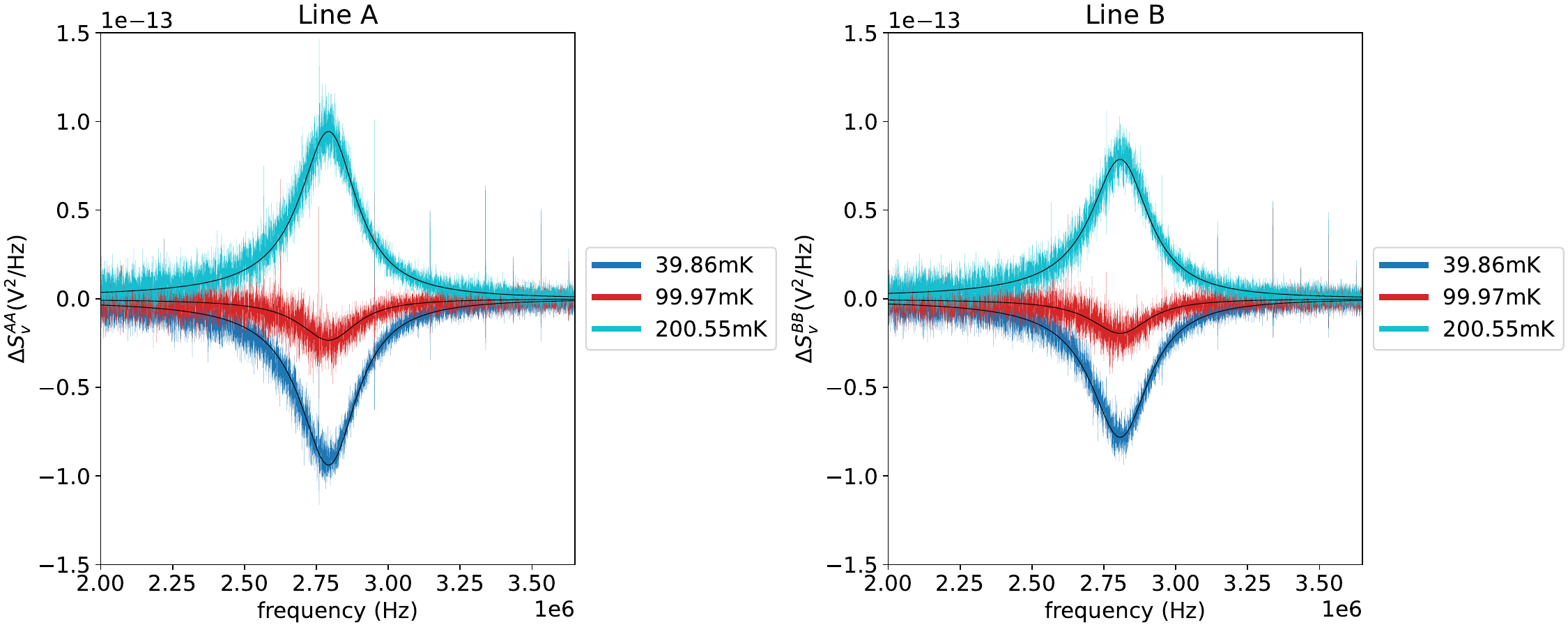}
    \caption{Differential noise spectra for line A (left) and B (right), corresponding to the raw spectra shown in Supplemental Fig.~\ref{fig:calib_raw}. The black lines are fits using Supplemental Eq.~\ref{eq:calib_thermalnoise}. We again show three typical sets of data and fit for the sake of visibility. Applying a smoothing filter to the data gives an estimation error smaller than $5\%$ of the parameters.}
    \label{fig:calib_diff}
\end{figure*}

\begin{figure*}[h!]
    \centering
    \includegraphics[width=0.97\textwidth]{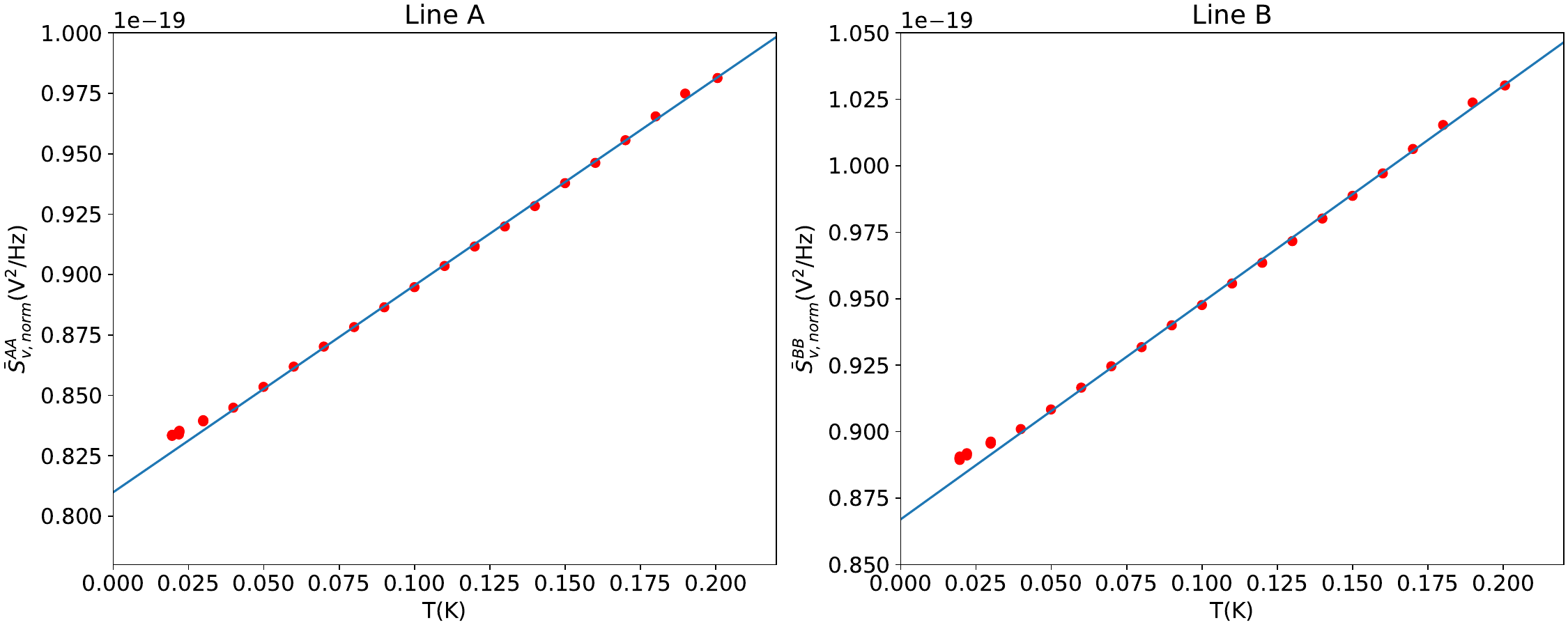}
    \caption{Integrated noise spectra from Supplemental Fig.~\ref{fig:calib_raw} versus temperature for $\nu=2$. The line is a linear fit of the data. $\bar{S}_{v,norm}^{AA}(T=0) = (8.099 \pm 0.007) \times 10^{-20}\, \mathrm{V}^2/\mathrm{Hz}$, $\bar{S}_{v,norm}^{BB}(T=0) = (8.670 \pm 0.007) \times 10^{-20}\, \mathrm{V}^2/\mathrm{Hz}$. }
    \label{fig:calib_int}
\end{figure*}

The total voltage noise of the amplifier, including the part originating from back flowing current noise through the RLC circuit, can be found from the intercept in temperature dependence of the integrated noise before amplifier $\bar{S}_{v,norm}^{AA/BB}/G_{A/B}^2$, as shown in Supplemental Fig.~\ref{fig:calib_int}. Note that this plot makes apparent the fact that the resonator and sample temperature start differing from the fridge temperature below $50~$mK. At $\mathrm{T}=0\,\mathrm{K}$, we omit the thermal noise from sample and RLC circuit resonator, leaving only the contribution of the amplifier. Therefore, performing temperature calibrations at different filling factors in the sample allows separating the current and voltage noise generated by the amplifier. Since the integrated noise before amplification is written as:
\begin{align}
    \bar{S}_{v,norm}^{AA/BB}(T=0) &= \frac{1}{G_{A/B}^2 \Delta f }\int_{\Delta f} df \frac{S^{AA/BB}_v(T=0)}{F(\frac{f}{f_cA/B}) } \\ \nonumber
    & = \ S_{v,amp}^{A/B} + \frac{\int_{\Delta f} df \left\lvert Z_{//}^{A/B}\right\rvert^2}{\Delta f} S^{A/B}_{i,amp} ,
    \label{eq:calib_int_norm}
\end{align} 
we obtain two equations for two variables $S_{v,amp}^{A/B}$ and $S^{A/B}_{i,amp}$ for $\nu=2$ and $\nu=3$. This yields an estimation of the amplifiers' voltage and current noises in Table \ref{tab:Svamp_Siamp}. Note that the finite current noise gives rise to a crossover between voltage and current noise at the resonance, roughly proportional to the imaginary part of the RLC impedance~\cite{Richards1986}, inducing an asymmetry in the resonances shown in the raw data shown Supplemental Fig. \ref{fig:calib_raw}. 

 Three separate calibration runs were performed over the two months during which the sample was measured. For each run, we have performed the calibration at bulk filling factor $\nu_\mathrm{EW}=2$ and $3$. The overall uncertainty on the value of the gain is about $6~\%$. Importantly, the extracted values are very similar, within the same uncertainty, to the values reported in previous noise measurements in graphene samples done in our group, focusing on the quantitative measurement of the quantized heat flow in QH edge channels~\cite{LeBreton2022,Delagrange2024}. Given this, the increased shot noise reported in the main text cannot be attributed to a measurement or calibration error.

\begin{table}
\begin{tabular}{ | m{2.0cm} | m{4.0cm}| m{4.0cm} |}
\hline
Amplifier  & $S_{v,amp}\,(\mathrm{V}^2/\mathrm{Hz})$ & $S_{i,amp}\, (\mathrm{A}^2/\mathrm{Hz})$\\
\hline
A & $(7.595\pm 0.028)\times 10^{-20}$ & $(7.382\pm 0.313)\times 10^{-28}$\\
\hline
B & $(8.432\pm 0.029)\times 10^{-20}$ & $(3.713\pm 0.337)\times 10^{-28}$\\
\hline
\end{tabular}
\caption{Amplifier voltage and current noise for line A and B.}
\label{tab:Svamp_Siamp}
\end{table}



\begin{table}[h!]
\begin{tabular}{ | m{0.5cm} | m{1.5cm}| m{1.5cm} |m{1.5cm} | m{1.5cm} | m{1.5cm} | m{1.5cm} | m{1.5cm} | m{1.5cm} |}
    
  \hline
    & A \newline $\nu=2$ \newline 07/07/22& B \newline $\nu=2$ \newline 07/07/22& A \newline $\nu=2$ \newline 08/04/22& B \newline $\nu=2$ \newline 08/04/22& A \newline $\nu=2$ \newline 08/30/22& B \newline $\nu=2$ \newline 08/30/22 & A \newline $\nu=3$ \newline 08/30/22& B \newline $\nu=3$ \newline 08/30/22\\
  \hline
  $R$ & 500 k$\Omega$ & 450 k$\Omega$ & 500 k$\Omega$ & 450 k$\Omega$ & 500 k$\Omega$ & 450 k$\Omega$ & 500 k$\Omega$ & 450 k$\Omega$\\
  \hline
  $L$ & 22.8 $\mu$H & 22 $\mu$H & 22.9 $\mu$H & 22.05 $\mu$H & 22.8 $\mu$H & 22.1 $\mu$H & 22.9 $\mu$H & 22.1 $\mu$H\\
  \hline
  $c_2$ & 143 pF & 145.2 pF & 147.2 pF & 145 pF & 142 pF & 145 pF & 142 pF & 145 pF\\
  \hline
  $r_L$ & 22 $\Omega$ & 22 $\Omega$ & 22 $\Omega$ & 22 $\Omega$ & 22 $\Omega$ & 22 $\Omega$ & 22 $\Omega$ & 22 $\Omega$\\
  \hline
  $G$ & 3480 & 3280 & 3430 & 3240 & 3430 & 3200 & 3450 & 3200 \\
  \hline
\end{tabular}
\caption{RLC circuit resonator parameters and overall gain extracted from the three temperature calibration runs.}
\label{tab:calib_result}
\end{table}

\section{Additional data at $B=13~$T}

\subsection{Conductance}
\begin{figure}[ht]
\centering

\includegraphics[width=0.45\textwidth]{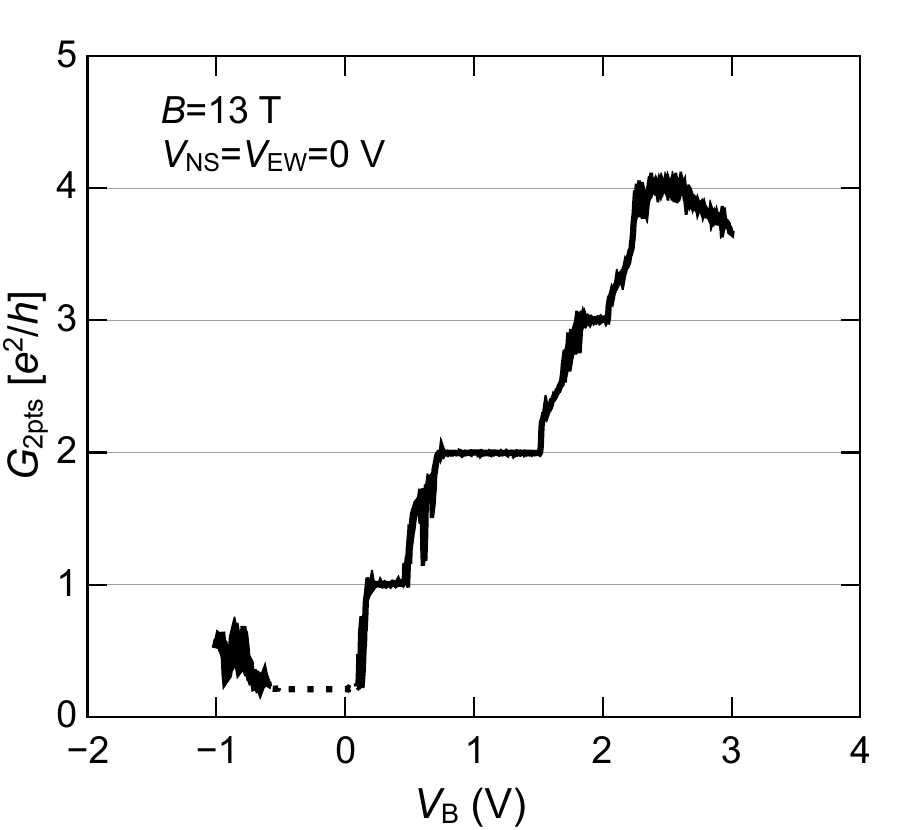}
\caption{\label{FigSup-GvsVB} 
Hall conductance (measured in a 2 point configuration) of the sample versus back gate voltage, at $B=13~$T and $T=100~$mK, for $V_\mathrm{EW}=V_\mathrm{NS}=0$. The dashed line region corresponds to $\nu=0$, where the lock-in preamplifiers saturate.
} 
\end{figure}
The QPC is operated at $B=13~$T by tuning the EW regions to $\nu=2$ using both back and EW gates. Supplemental Fig.~\ref{FigSup-GvsVB} shows a 2 point measurement of the Hall conductance versus back gate voltage, displaying the plateaus for $\nu=0$, $1$, $2$, $3$ and $4$. We set the the value of the back gate voltage to the $\nu=2$ plateau, then sweep the EW gate voltage to locate the range of the $\nu=0$ plateau that will be used for the NS QPC gate. Having fixed both the back gate and the NS gate, we sweep the EW gate along the $\nu=2$ plateau to obtain the conductance data shown in main text Fig.~1.

\begin{figure}[ht]
\centering
\includegraphics[width=0.45\textwidth]{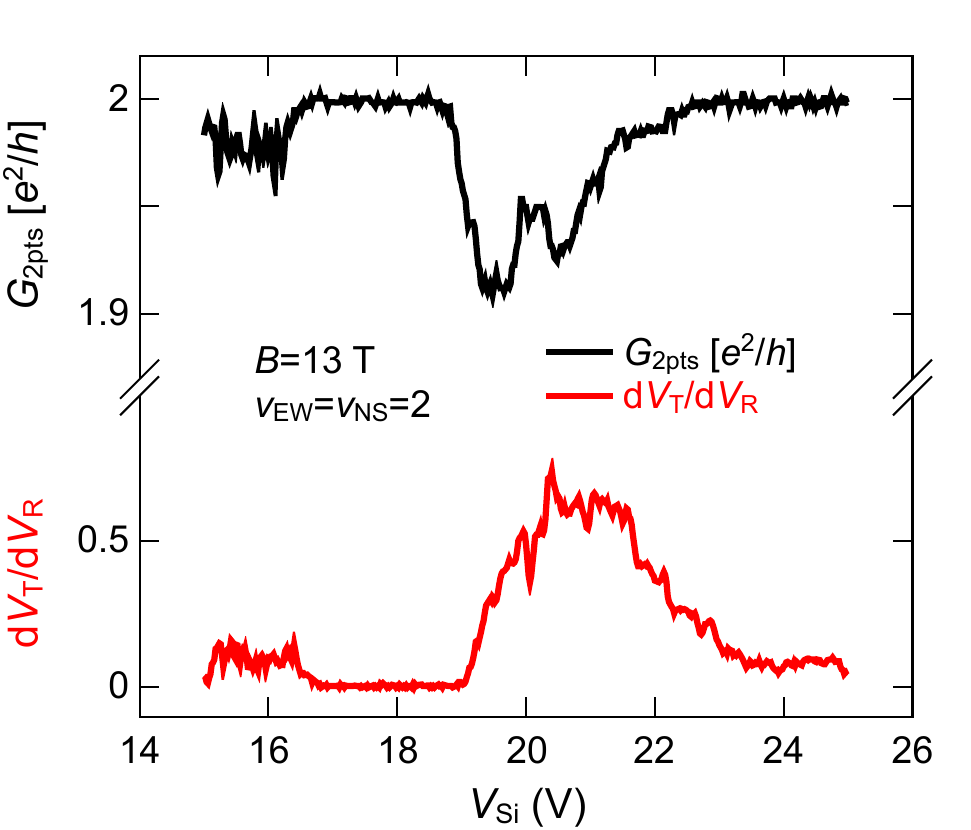}
\caption{\label{FigSup-GvsVsi} 
Hall conductance (measured in a 2 point configuration - black) and transconductance $\mathrm{d}V_\mathrm{T}/\mathrm{d}V_\mathrm{R}$ (red) versus silicon back gate voltage, for bulk filling factors $\nu_\mathrm{EW}=\nu_\mathrm{NS}=2$, at $B=13~$T.
}
\end{figure}

In addition to the graphite gate, the silicon back gate is used to locally dope the graphene flake close to its metallic edge contacts. The voltage applied to the silicon gate is selected by measuring the variation of the transconductance signal $\mathrm{d}V_\mathrm{T}/\mathrm{d}V_\mathrm{R}$, which reflects the quality of the grounded contact downstream of contact R. Supplemental Fig.~\ref{FigSup-GvsVsi} shows the result for $\nu_\mathrm{EW}=\nu_\mathrm{NS}=2$, at $B=13~$T: around $V_\mathrm{Si}=18~$V, the signal vanishes, signaling an optimal transmission to the grounded contact. In the results shown in the main text we have thus set $V_\mathrm{Si}=18~$V.

\begin{figure}[ht]
\centering
\includegraphics[width=0.45\textwidth]{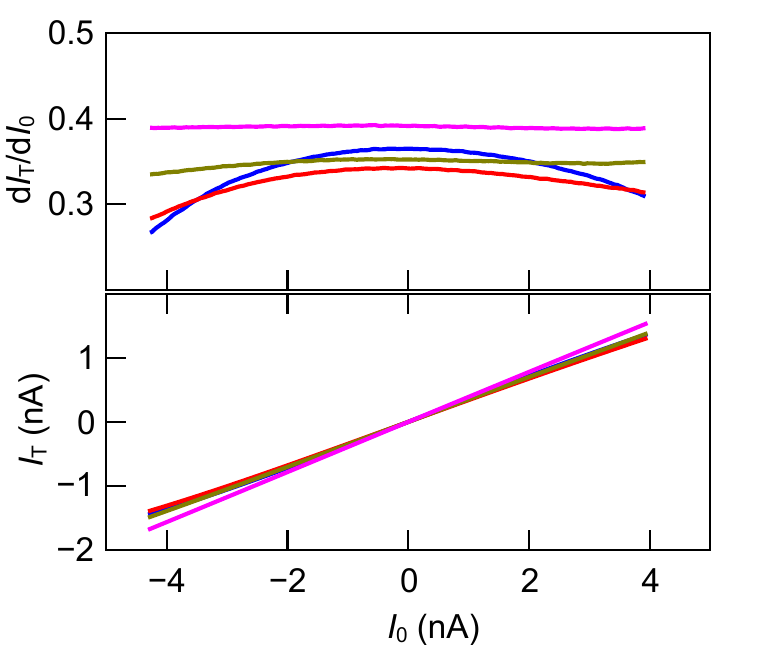}
\caption{\label{FigSup-GITvsItot}  Top: QPC transmission $\mathrm{d}I_\mathrm{T}/\mathrm{d}I_\mathrm{0}$ versus dc current $I_0$, for different gate voltages corresponding to the transmission of the outer channel. Bottom: corresponding transmitted current $I_\mathrm{T}$ obtained by integration.
}
\end{figure}

The dc current transmitted across the QPC $I_\mathrm{T}$ is obtained by measuring the differential transmission $\mathrm{d}I_\mathrm{T}/\mathrm{d}I_\mathrm{0}$ at finite dc current $I_0$, which is then integrated. Supplemental Fig.~\ref{FigSup-GITvsItot} shows typical examples of this procedure, for several EW gate voltages. The average QPC conductance $\bar{G}_\mathrm{QPC}$ is directly obtained from the $\mathrm{d}I_\mathrm{T}/\mathrm{d}I_\mathrm{0}$ measurements.

\newpage

\subsection{Noise}
\subsubsection{Extracting the noise from auto- and cross-correlations}

\begin{figure}[ht]
\centering
\includegraphics[width=0.45\textwidth]{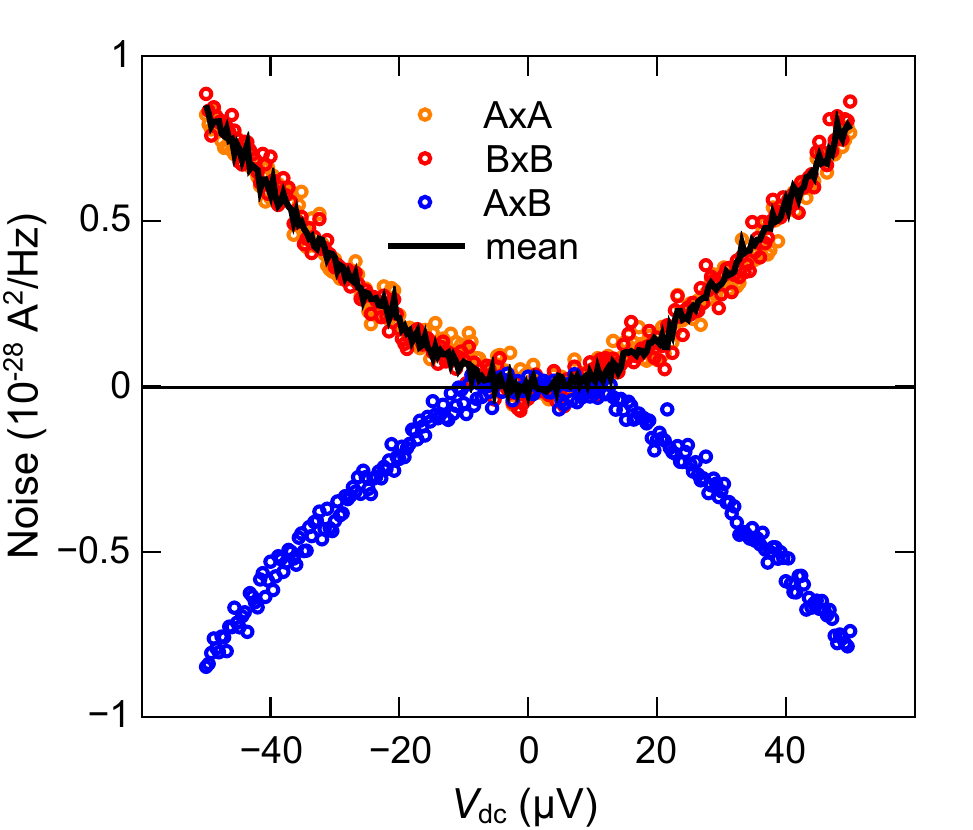}
\caption{\label{FigSup-autovscross}  Auto-correlated (orange and red symbols) and cross-correlated (blue symbols) noises, as well as their weighted average (black line, see text) versus $\Vdc$, at $T=100~$mK, for $\tau_1=0.86$.
}
\end{figure}

According to Supplemental Eq.~\ref{eq:calib_thermalnoise2}, the shot noise generated by the sample can be obtained from three independent measurements: the auto-correlations from each noise measurement line, and their cross-correlation. Comparing these signals allows discarding additional sources of noise, such as noise added at the contacts before or after the QPC. Indeed, in absence of these sources, both auto-correlations should be equal, and the cross-correlation exactly opposite to the auto-correlations. Differing auto-correlations imply that there is an additional source of noise between the QPC and one of the noise measurement contacts, most likely due to one of these contacts being improperly coupled to the edge channels. Equal auto-correlations with a smaller (in absolute value) cross-correlation implies that there is an additional noise source between the current feed contact and QPC. Indeed, in the auto-correlations, both this additional noise and the QPC shot noise add incoherently in absolute value, while the cross-correlation is given by their difference (thus, in absence of the additional noise, the cross-correlation is exactly equal to minus the thermal noise). We can thus extract the QPC shot noise by computing the weighted average between auto-correlation and cross-correlation $((A\times A+B \times B)/2-A\times B)2$. Systematically calculating this average (provided of course that the autocorrelations are equal) has the added benefit of increasing our signal-to-noise ratio. In the main text, all the noise data are extracted from the computed average. We show examples of the procedure in Fig.~\ref{FigSup-autovscross}. The auto-correlations are shown as red and orange symbols, the cross-correlation as blue symbols, and the average as a black line. We obeserved that there was no added noise before or after the QPC, thus confirming that the enhanced noise stems from scattering at the QPC.

\subsubsection{Scattering formalism expression of the noise versus transmitted current}
The noise measurements are fitted with main text Eq.~1, which expresses the noise as a function of the transmitted current $I_\mathrm{T}$, with the $I_\mathrm{T}$ and $\bar{G}_\mathrm{QPC}$ obtained from conductance measurements (see above), the temperature fixed to $T=100~$mK, and the Fano factor $F$ as the only free parameter. Main text Eq.~1 has a slightly unusual expression, as the noise is more commonly expressed as a function of the dc voltage $\Vdc$ applied to the sample~\cite{Blanter2000}:

\begin{equation}
\Delta S (\Vdc) = \frac{2 e^2}{h} \sum_i \tau_i (1-\tau_i) \left(  e \Vdc \times \mathrm{coth}(\frac{e \Vdc}{2 \kB T}) -2 \kB T \right).
\label{eq:shotnoiseVdc}
\end{equation}

This equation can be expressed as a function of the transmitted current $I_\mathrm{T}=\bar{G}_\mathrm{QPC}\Vdc$:

\begin{equation}
\Delta S (I_\mathrm{T}) = \frac{2 e^2}{h} \sum_i \tau_i (1-\tau_i) \left(  e \frac{I_\mathrm{T}}{\bar{G}_\mathrm{QPC}} \times \mathrm{coth}(\frac{e I_\mathrm{T}}{2 \bar{G}_\mathrm{QPC}\kB T}) -2 \kB T \right).
\label{eq:shotnoiseIT}
\end{equation}

The above equation thus already contains the $I_\mathrm{T}\times\mathrm{coth}(\frac{e I_\mathrm{T}}{2 \bar{G}_\mathrm{QPC}\kB T})$ of main text Eq.~1, and can be further rewritten as:

\begin{equation}
\Delta S (I_\mathrm{T}) = 2\frac{e^2}{h}\frac{e}{\bar{G}_\mathrm{QPC}} \sum_i \tau_i (1-\tau_i) \left(  I_\mathrm{T} \times \mathrm{coth}(\frac{e I_\mathrm{T}}{2 \bar{G}_\mathrm{QPC}\kB T}) -2\bar{G}_\mathrm{QPC}\frac{\kB T}{e} \right).
\label{eq:shotnoiseIT2}
\end{equation}

Recalling that $\bar{G}_\mathrm{QPC}=\frac{e^2}{h}\sum_i \tau_i$, we obtain:

\begin{equation}
\Delta S (I_\mathrm{T}) = 2e \frac{\sum_i \tau_i (1-\tau_i)}{\sum_i \tau_i } \left(  I_\mathrm{T} \times \mathrm{coth}(\frac{e I_\mathrm{T}}{2 \bar{G}_\mathrm{QPC}\kB T}) -2\bar{G}_\mathrm{QPC}\frac{\kB T}{e} \right).
\label{eq:shotnoiseIT3}
\end{equation}.

Using the standard definition of the Fano factor~\cite{Blanter2000} $F=\frac{\sum_i \tau_i (1-\tau_i)}{\sum_i \tau_i }$, we finally obtain main text Eq.~1:

\begin{equation}
\Delta S (I_\mathrm{T}) = 2 e F \left(  I_\mathrm{T} \times \mathrm{coth}(\frac{e I_\mathrm{T}}{2 \bar{G}_\mathrm{QPC} \kB T}) -2 \bar{G}_\mathrm{QPC} \frac{\kB T}{e} \right),
\label{eq:shotnoise}
\end{equation}

\subsubsection{Noise fits with a prefactor on $I_\mathrm{T}$}

\begin{figure}[ht]
\centering
\includegraphics[width=0.45\textwidth]{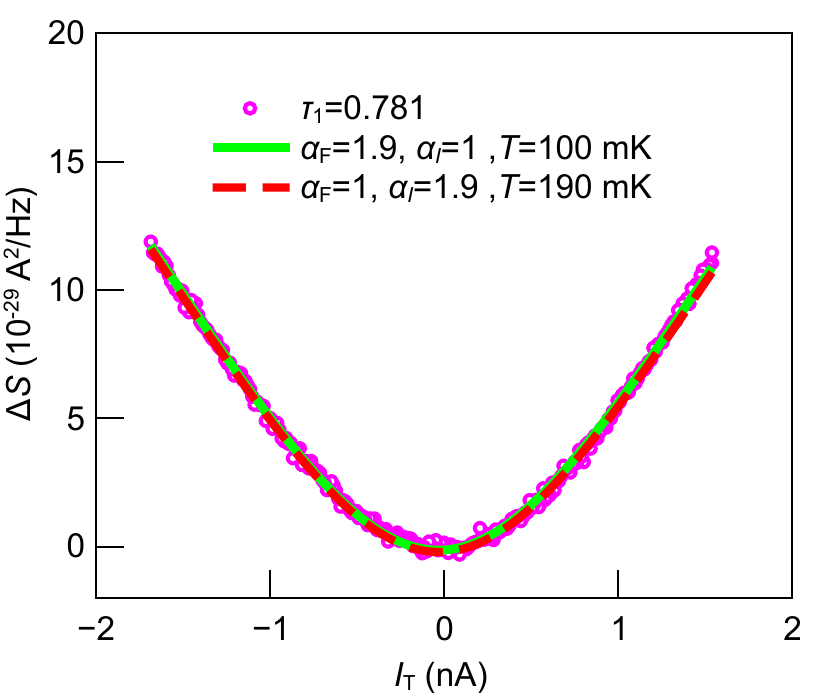}
\caption{\label{FigSup-NoisevsI-fitsdiffT} 
Symbols: noise versus transmitted current, for $\tau_1=0.781$. Green continuous line: fit with the Fano factor prefactor $\alpha_\mathrm{F}$ used as a fit parameters, with the temperature fixed to $100~$mK and no prefactor on the transmitted current. Red dashed line: fit with a fixed $\alpha_\mathrm{F}=1$, and the temperature and the transmitted current prefactor used as fit parameters. 
}
\end{figure}

Besides the calibration of the noise measurement setup, the calibration of the transmitted dc current can affect the values of the extracted Fano factors. If our results were due to an error on the dc current, this error would have to be more than $50~\%$. The line through which we apply the dc current $I_0$ is the same through which we apply the lock in excitation $\mathrm{d}I_0$ (see Supplemental Fig.~\ref{figsup-conductances}). The quantized values of the differential conductances measured at a few Hz that are shown in the main text attest that the calibration of this line is correct. To further discard this hypothesis, we have also fitted the noise versus $I_\mathrm{T}$ data shown in main text Fig.~2 with a modified version of main text Eq.~1, where a prefactor $\alpha_I$ in front of the transmitted current is used as a fit parameter instead of the prefactor $\alpha$ that use for the Fano factor (thus, $\alpha$ is fixed to 1). The only way to obtain reasonable fits is then to also let the temperature be a fit parameter, allowing to reproduce the thermal rounding at low bias. Supplemental Fig.~\ref{FigSup-NoisevsI-fitsdiffT} shows such a fit, along with that used in the main text: a good fit of the data can be obtain with $\alpha=1$, at the cost of having a large $\alpha_I=1.9$ as well as a large temperature $T=190~$mK. This is doubly unlikely, as not only this would imply a large error on the current calibration, but also one on the electron temperature. From the calibration of the noise measurement, and also from previous experiments in the same fridge showing base electron temperatures as low as $11~$mK~\cite{LeBreton2022}, we know that the electron temperature in the sample closely follows that of our refrigerator above $60~$mK.

\subsubsection{Quadratic noise fits}

\begin{figure}[ht]
\centering
\includegraphics[width=0.85\textwidth]{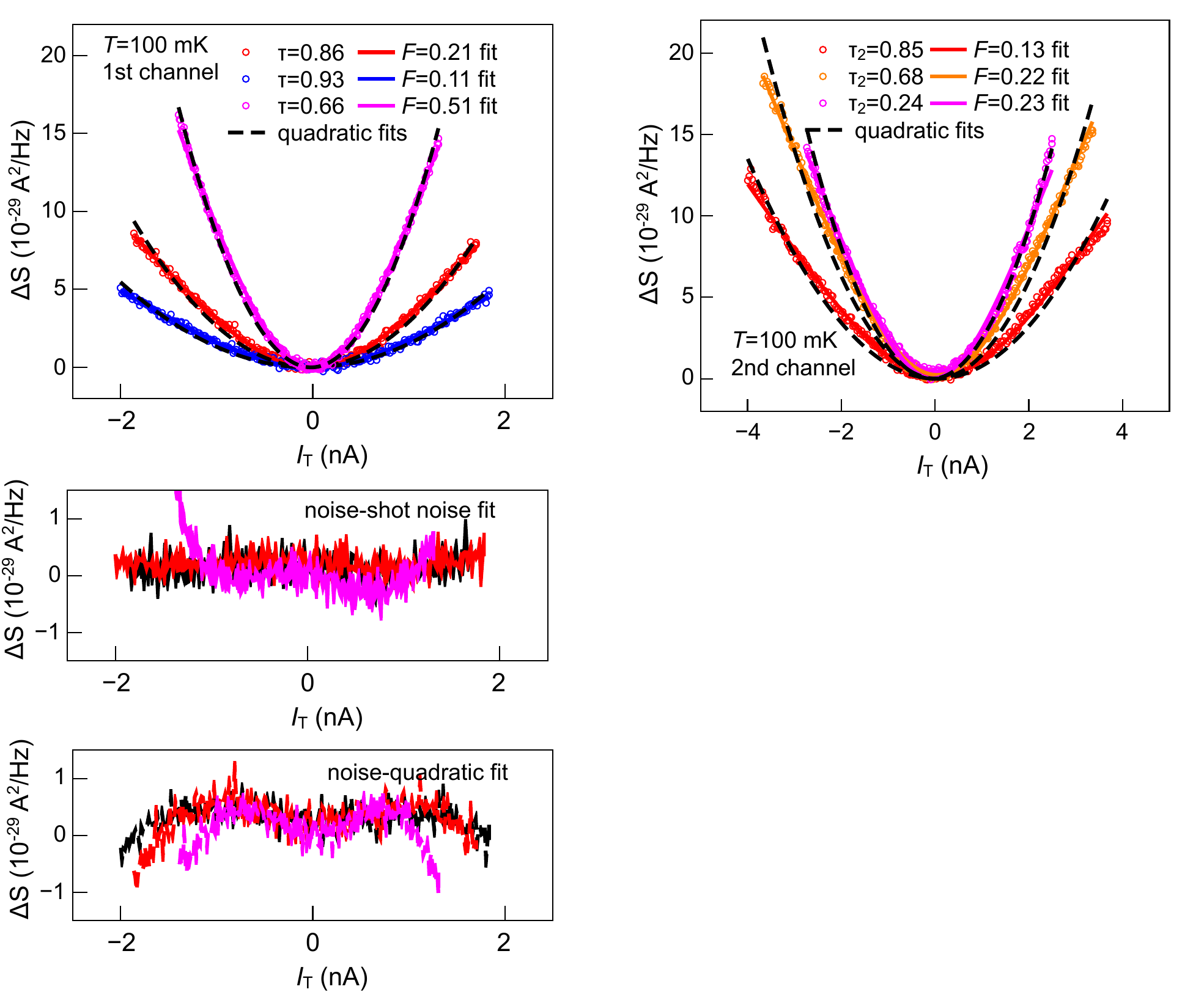}
\caption{\label{FigSup-quadraticfits} Quadratic fits of typical noise versus $I_\mathrm{T}$ data, for the outer edge channel (left) and the inner edge channel (right). The symbols are experimental data, with colors corresponding to different settings of the QPC transmissions. The colored full lines are the shot noise fits of main text Eq.~1. The dashed lines are quadratic fits of the data. The bottom left plots are the difference between the outer channel data and the shot noise (full lines) and quadratic (dashed lines) fits.
}
\end{figure}

Our observed increased in the shot noise could be attributed to $1/f$ noise increasing in the sample at finite bias. The typical signatures of such an increase are noisy conductance data at finite bias, and a quadratic dependence of the noise with the bias. Supplemental Fig.~\ref{FigSup-quadraticfits} shows a comparison between the shot noise fits of main text Eq.~1 and quadratic fits for several typical datasets of both channels. The shot noise fits are generally better than the quadratic fits, which is emphasized by plotting the difference between the noise and the fits (bottom plots in supplemental Fig.~\ref{FigSup-quadraticfits}). In addition, the conductance data, for instance in the plots of main text Fig. 1, do not show any significant increase in the fluctuation, precluding a noise increase due to $1/f$ noise.

\subsubsection{Definition of the Fano factor with respect to the QPC or the quantuml dot}

As plotted in main text Fig.~3a), the Fano factor corresponds to the ratio between the excess noise and the total current transmitted through the QPC. Plotting it as in main text Fig.~3a) versus conductance across the QPC thus corresponds to the usual plot for a QPC $V$ versus $\sum \tau_i$, where $\tau_i$ is the transmission of the $i$-th channel across the QPC. There are however several other ways to plot the same noise data using different definition of the Fano factor: individual Fano factors for each channel, and Fano factor defined in reflection rather than in transmission. Recalling that $I_0$ is the total dc current flowing towards the QPC, with each channel carrying $I_0/2$, one can define the transmitted and reflected currents for each channel through the QPC and or the quantum dot:

\begin{multline}\\
    I^{(1)}_\mathrm{T,QPC}=\tau_1 I_0/2\\
    I^{(1)}_\mathrm{R,QPC}=(1-\tau_1) I_0/2\\
    I^{(2)}_\mathrm{T,QPC}=\tau_2 I_0/2\\
    I^{(2)}_\mathrm{R,QPC}=(1-\tau_2) I_0/2\\
    I^{(1)}_\mathrm{T,QD}=(1-\tau_1) I_0/2=I^{(1)}_\mathrm{R,QPC}\\
    I^{(1)}_\mathrm{R,QD}=\tau_1 I_0/2=I^{(1)}_\mathrm{T,QPC}\\
    I^{(2)}_\mathrm{T,QD}=\tau_2 I_0/2=I^{(2)}_\mathrm{T,QPC}\\
    I^{(2)}_\mathrm{R,QD}=(1-\tau_2) I_0/2=I^{(2)}_\mathrm{T,QPC}\\
    \label{eq-trans-refl-currents}
\end{multline}

The inversion between the current transmitted through the QPC and through the quantum dot for the outer channel (channel $i=1$) again stems from the fact that the quantum dot couples the outer edge channel from the west side to the east side in reflection (see main text). Thus, one can define 4 different channel-specific Fano factors:

\begin{multline}\\
    F^{(1)}_\mathrm{T,QPC}=F^{(1)}_\mathrm{R,QD}=\frac{\Delta S}{2 e I^{(1)}_\mathrm{T,QPC}}=\frac{\Delta S}{2 e I^{(1)}_\mathrm{R,QD}}=\frac{\Delta S}{2 e \tau_1 I_0/2}\\
    F^{(1)}_\mathrm{R,QPC}=F^{(1)}_\mathrm{T,QD}=\frac{\Delta S}{2 e I^{(1)}_\mathrm{R,QPC}}=\frac{\Delta S}{2 e (1-\tau_1) I_0/2}=F^{(1)}_\mathrm{T,QPC}\times\frac{\tau_1}{(1-\tau_1)}\\
    F^{(2)}_\mathrm{T,QPC}=F^{(2)}_\mathrm{T,QD}=\frac{\Delta S}{2 e I^{(2)}_\mathrm{T,QPC}}=\frac{\Delta S}{2 e I^{(2)}_\mathrm{T,QD}}=\frac{\Delta S}{2 e \tau_2 I_0/2}\\
    F^{(2)}_\mathrm{R,QPC}=F^{(1)}_\mathrm{R,QD}=\frac{\Delta S}{2 e I^{(2)}_\mathrm{R,QPC}}=\frac{\Delta S}{2 e (1-\tau_2) I_0/2}=F^{(2)}_\mathrm{T,QPC}\times\frac{\tau_2}{(1-\tau_2)}\\
    \label{eq-trans-refl-Fanos}
\end{multline}

\begin{figure}[ht]
\centering
\includegraphics[width=0.75\textwidth]{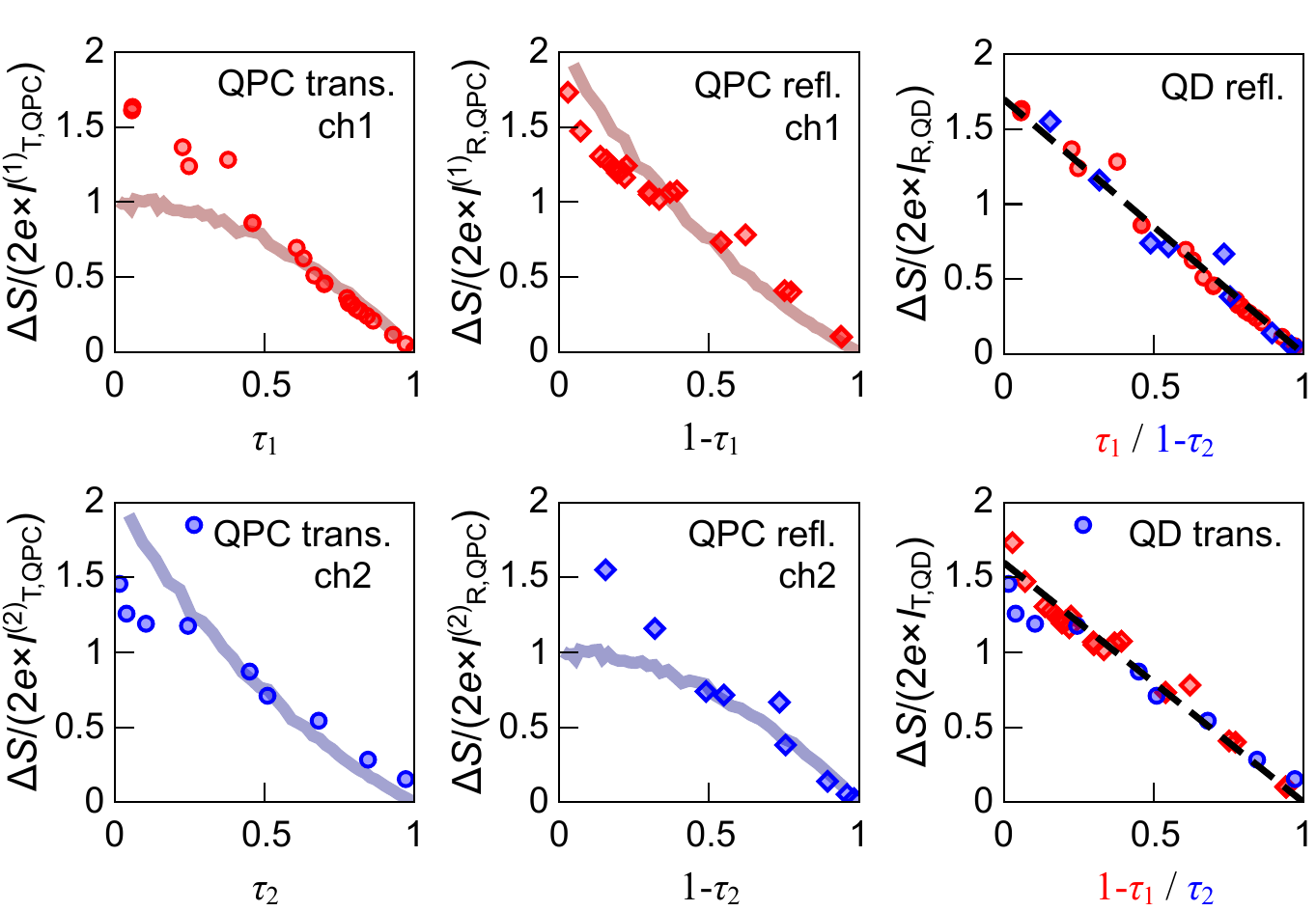}
\caption{\label{FigSup-Fanovstau-4configs} Fano factor for each channel (red: outer, blue: inner), corresponding to the dataset of main text Fig.~3a), defined and plotted with respect to the QPC transmission (left column) and reflection (middle column). The thick line is the result of the simplified model for a single channel. Right column: Fano factor for each channel defined and plotted with respect to the quantum dot reflection (upper right panel) and transmission (lower right channel). The black dashed lines are linear functions.
}
\end{figure}

Using the data shown in main text Fig.~3a), we define $\tau_1$ and $\tau_2$ from the measured $\bar{G}_\mathrm{QPC}$, allowing to plot the four above-defined Fano factors versus their respective transmission/reflection. Supplementary Fig.~\ref{FigSup-Fanovstau-4configs} shows the corresponding plots, generalizing and confirming for all configurations the observation made in the main text. The Fano factor shows a linear decrease with transmission/reflection, with similar quantitative behaviors when plotting the Fano factor with respect to the quantum dot. The black dashed lines in the right panels are functions $\alpha(1-x)$, with a prefactor $\alpha=1.7$ for the quantum dot reflection data (upper right panel), and $\alpha=1.6$ for the quantum dot transmission data (lower right panel). We also plot the results of the simplified model (thick lines in the left and middle columns): when plotted in reflection with respect to the quantum dot (upper right and lower middle panels), the calculated Fano factor saturates to 1 at low quantum dot reflection, again showcasing the fact that since this model does not describe the internal dynamics of the dot, when most of the charges go through the quantum dot, rare reflection events can only involve a single charge.

\subsubsection{Noise measurements at 20~mK}

\begin{figure}[ht]
\centering
\includegraphics[width=0.85\textwidth]{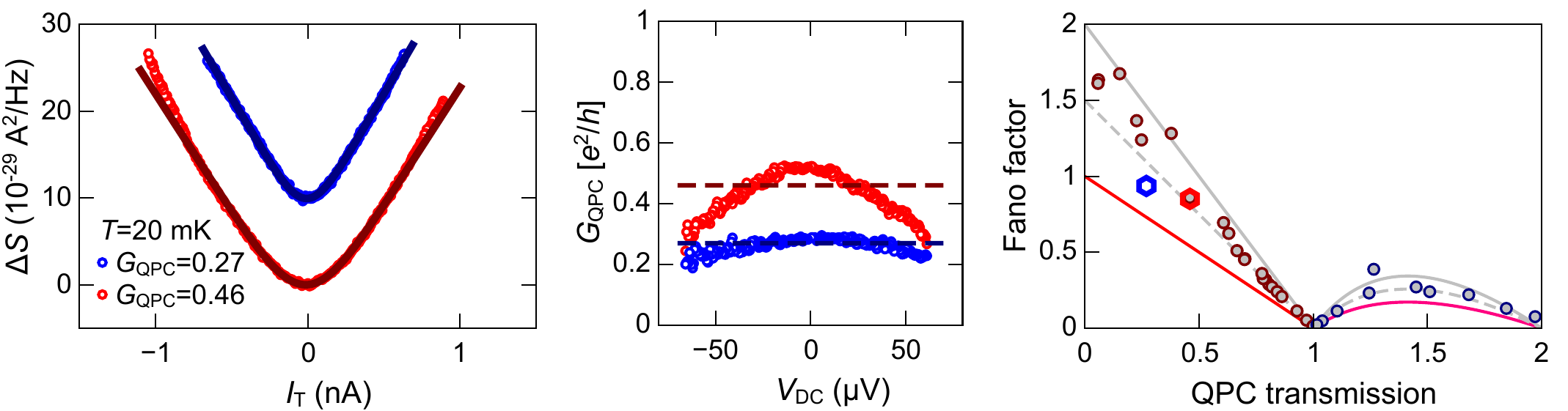}
\caption{\label{FigSup-data20mK} Noise measurements at $T=20~$mK. Left: excess noise versus $I_\mathrm{T}$, for two different settings of the QPC partially transmitting the outer edge channel. Symbols: experimental data, black lines: fits using main text Eq. 1, with the electron temperature $T\approx56~$mK as an additional fit parameter. Middle: corresponding conductance versus $\Vdc$ measurements. The dashed lines correspond to the average conductances used in the fits in the left panel. Right: Fano factor versus QPC transmission, corresponding to main text Fig.~3. The circles are the data obtained at 100~mK, and the red and blue hexagons to the Fano factors extracted from the fits of the 20~mK shown in the left panel.
}
\end{figure}

Supplemental Fig.~\ref{FigSup-data20mK} shows additional noise data taken at a fridge temperature of 20~mK, for two gate voltages settings of the first channel of the QPC (corresponding to a partial transmission of the outer edge channel). The noise data (left panel) is well matched by the fits of main text Eq.~1, albeit with a fitted electron temperature markedly higher than that of the fridge, $T\approx56~$mK. To avoid having to use the electron temperature as an additional fit parameter, we have thus opted to perform the majority of the noise measurements at 100~mK where the electron temperature matches well that of the fridge. Note that the data at 20~mK also shows an increased Fano factor, comparable with that of the 100~mK data, as shown in the right panel of Fig.~\ref{FigSup-data20mK}.

\subsubsection{Noise measurements at larger bias}

\begin{figure}[ht]
\centering
\includegraphics[width=0.85\textwidth]{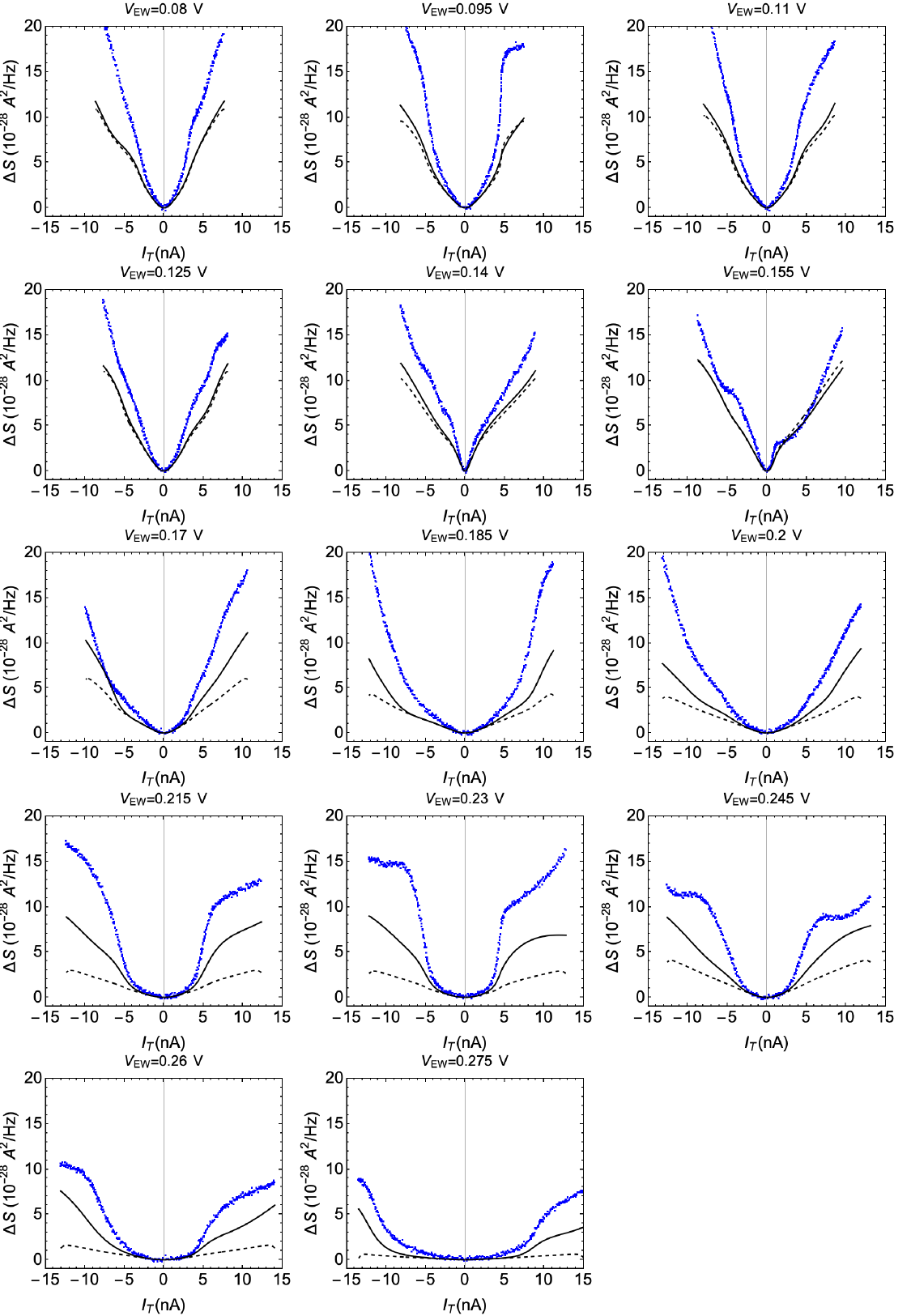}
\caption{\label{FigSup-noisevsItlargebias} Noise versus $I_\mathrm{T}$ corresponding to the data shown in main text Fig.~5a), for $V_\mathrm{EW}$ in the range of the low-energy noise data shown in main text Fig. 2. Blue dots: experimental data, black lines: scattering formalism calculations using the $\Vdc$-dependent conductance (full lines), and the low-bias average conductance (dashed lines).
}
\end{figure}

We show in Supplemental Fig.~\ref{FigSup-noisevsItlargebias} linecuts of the measured noise versus transmitted current, corresponding to the data shown in main text Fig.~5a) for the first channel, for various $V_\mathrm{EW}$ ranging from $0.08~$V to $0.275~$V (thus roughly corresponding to the gate voltage span of the low-energy noise data of main text Fig.~2). The data shows strong nonlinearities at large transmitted current, corresponding to the edge of the Coulomb diamonds. We plot at the same time the expected noise within a single-particle scattering formalism, using either main text Eq.~1 with an average value of the conductance calculated for $|\Vdc|<50~\mu$V (dashed lines), as well as the results of this formalism for energy-dependent transmission~\cite{Blanter2000}:

\begin{multline}
S (\Vdc) = \frac{2 e^2}{h} \int d\epsilon \times T(\epsilon)\left[f_L(\epsilon,\Vdc)(1-f_L(\epsilon,\Vdc)) + f_R(\epsilon,0)(1-f_R(\epsilon,0))\right]\\
+\frac{2 e^2}{h} \int d\epsilon\times T(\epsilon)(1-T(\epsilon))(f_L(\epsilon,\Vdc)-f_R(\epsilon,0))^2.
\label{eq:shotnoiseT(E)}
\end{multline}

The calculations show that the measured noise is systematically larger than the calculated noises (which reflects the large values of Fano factor reported in main text Fig.~5), and that the differences between the two calculations are small at low bias, and only become significant again at the edges of the Coulomb diamonds. The fact that even the energy-dependent calculation cannot account for the measured increased noise can be understood by noting that superpoissonian values of noise in quantum dots are linked to many-body effects, while scattering formalism calculations are single-particle.

 \clearpage
\section{Additional data at $B=6~$T}

\begin{figure}[ht]
\centering
\includegraphics[width=0.45\textwidth]{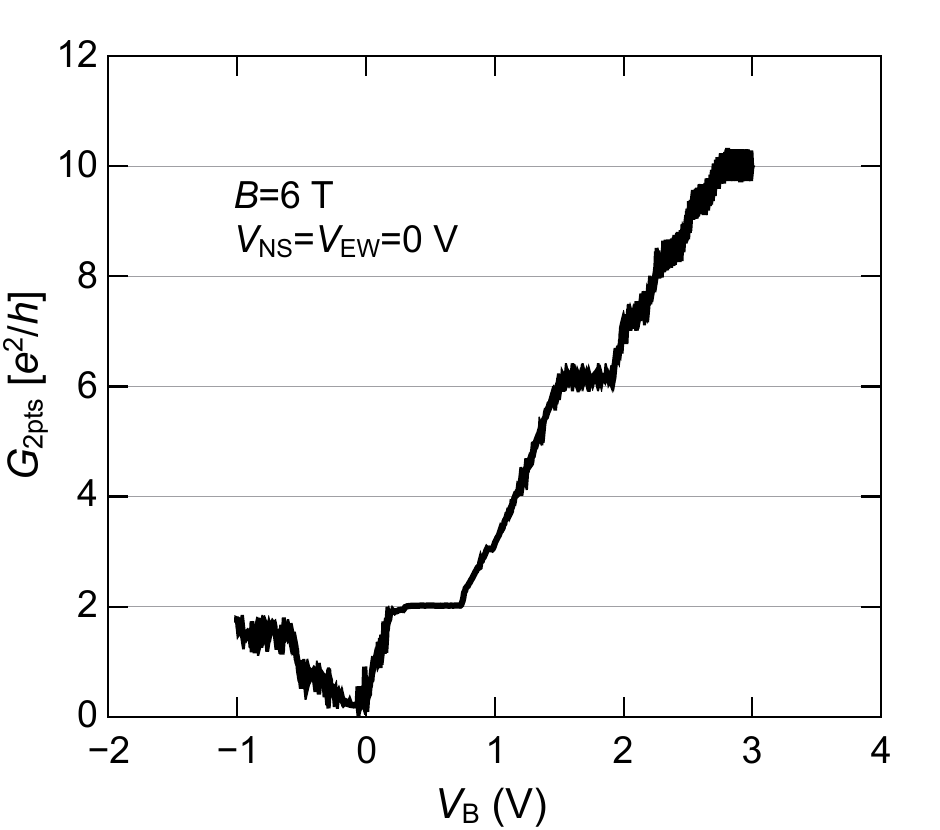}
\caption{\label{FigSup-GvsVB-6T} Hall conductance (measured in a 2 point configuration) of the sample versus back gate voltage, at $B=6~$T and $T=300~$mK, for $V_\mathrm{EW}=V_\mathrm{NS}=0$. 
}
\end{figure}

\begin{figure*}[ht]
\centering
\includegraphics[width=0.9\textwidth]{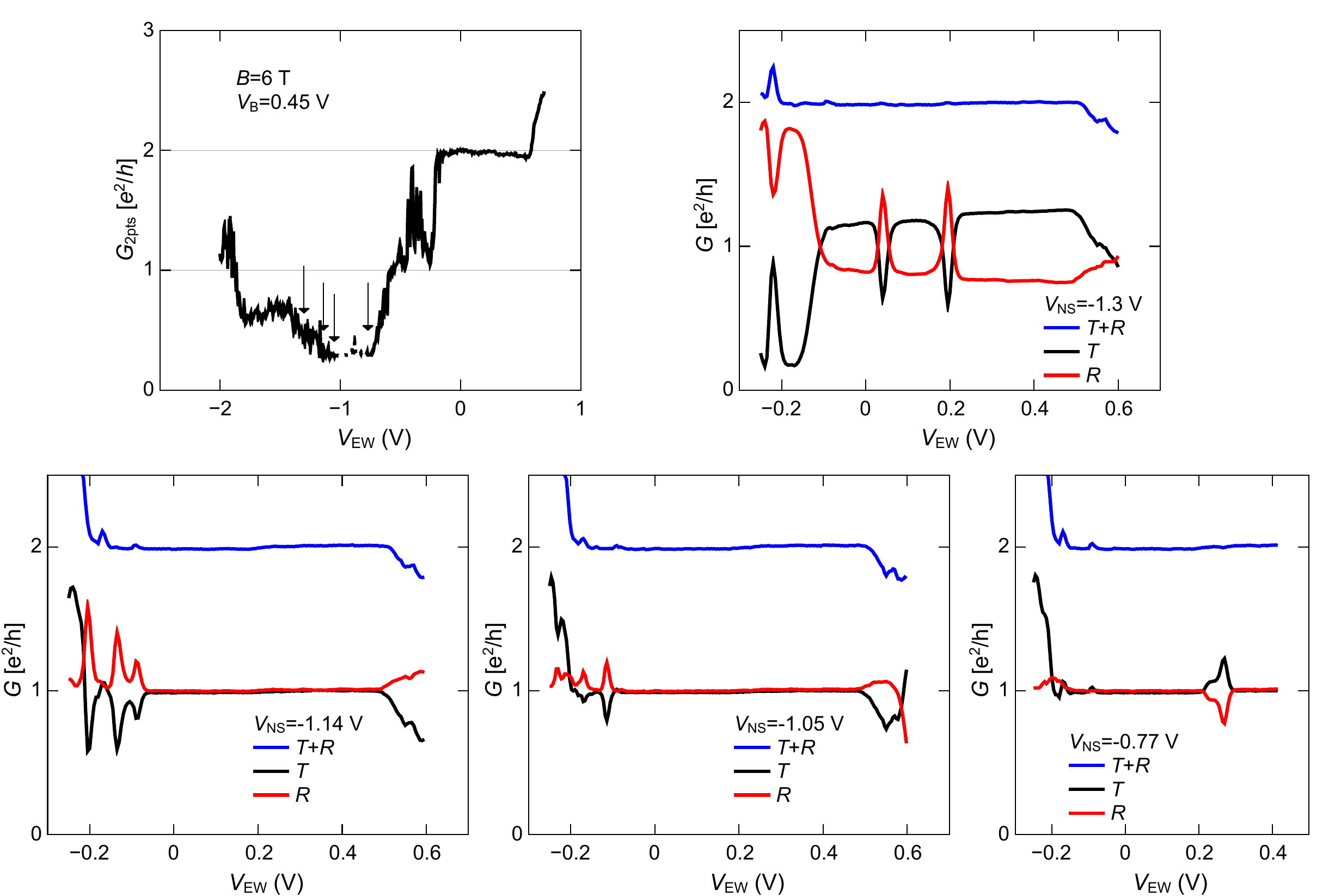}
\caption{\label{FigSup-QPCtraces-6T} Upper left: Hall conductance (measured in a 2 point configuration) of the sample versus $V_\mathrm{EW}$, at $B=6~$T, $T=300~$mK, and for $V_\mathrm{B}=0.45~$V. Upper right, and lower panels: differential conductances (similar to main text Fig.~2a: black (resp. red): transmitted (resp. reflected) conductance. Blue: sum of the transmitted and reflected conductances) versus $V_\mathrm{EW}$, for various values of $V_\mathrm{NW}$ indicated by the arrows in the upper left panel, corresponding to $\nu_\mathrm{NS}\approx0$.
}
\end{figure*}

 \clearpage
\section{Plasmon model for the enhancement of the Fano factor through tunneling across a quantum Hall island at $\nu=2/3$}

\subsection{Model}
We consider two integer quantum Hall channels at filling factor $\nu=1(+1)$ (were the additional $\nu=1$ channel in parenthesis can be seen as an inert plateau not involved in the dynamics) playing the role of leads and coupled by means of a fractional quantum Hall dot at $\nu=2/3(+1)$ (emerging again on top of the same inert integer plateau). The electron tunneling between these leads (indicated in the following as right $R$ and left $L$ respectively) and the dot will be considered as local.

In the following we will discuss the Hamiltonian contributions describing the various parts of this system.  

\subsubsection{Hamiltonians of the leads}
The leads are described by the Hamiltonians (from now one we will consider $\hbar=1$) 
\begin{equation}
\hat{H}_{R/L}=\mp i v \int^{+\infty}_{0} dx \hat{\Psi}_{R/L}^{\dagger}(x) \partial_{x} \hat{\Psi}_{R/L}(x)
\label{H_l}
\end{equation}
with $\hat{\Psi}_{R/L}(x)$ the annihilation field operators for an electron in the $R$ and $L$ reservoir respectively. 
In the above Equation we have considered, without loss of generality, the fact that the two leads can be described in terms of one-dimensional chiral free fermions propagating with velocity $v$ along semi-infinite edge channels. According to this, the reservoirs at fixed temperature $T$ and voltages $V_{L/R}$ can be described in terms of the Fermi distributions 

\begin{equation}
f_{L/R}(E)=\frac{1}{1+ \exp{\left[\left(E-e V_{L/R}\right)/k_{B}T \right]}}
\label{f_l}
\end{equation}
where $e$ is the electron charge and $k_{B}$ is the Boltzmann constant.   

In order to drive the system out of equilibrium and induce a current across the dot, a constant voltage bias 
\begin{equation}
V=V_{L}-V_{R}
\end{equation}
is applied between the reservoirs. For sake of simplicity we will assume the voltage drops as symmetric.

\subsubsection{Hamiltonian of the dot}

Recalling the conventional picture for the edge channels of a quantum Hall state at $\nu=2/3$, the Hamiltonian of a quantum dot of circumference $L$ can be written as~\cite{Johnson1991, Kane1994, Kane1995, Park2015}
\begin{eqnarray}
\hat{H}_{D}&=&\frac{1}{4 \pi}\int^{L}_{0} dx \left[v_{1} \left(\partial_{x} \hat{\phi}_{1} \right)^{2}+3 v_{2} \left(\partial_{x} \hat{\phi}_{2} \right)^{2}+2 v_{12}\partial \hat{\phi}_{1} \partial \hat{\phi}_{2}\right]\nonumber\\
&+& \int^{L}_{0} dx \left[\xi(x) e^{i \left(\hat{\phi}_{1}+3 \hat{\phi}_{2} \right)}+H.c. \right],
\label{H_D_12}
\end{eqnarray}
where we have omitted the dependence of the fields on the variable $x$ for notational convenience. In the above Equation the bosonic field $\hat{\phi}_{1}(x)$ describes the charge density fluctuations of an edge state at filling factor $\nu_{1}=1$ propagating with velocity $v_{1}$, while the field $\hat{\phi}_{2}(x)$ the ones of an edge state at $\nu_{2}=-1/3$ (counter-propagating with respect to the previous one) with velocity $v_{2}$. 
These fields satisfy the commutation relations 

\begin{equation}
\left[\hat{\phi}_{j}(x), \hat{\phi}_{k}(y) \right]=i \pi \nu_{j}\delta_{j,k} \mathrm{sgn}(x-y)\,\,\,(j,k=1,2)
\end{equation}
and are coupled by means of a short-range interaction term with intensity $v_{12}$. Moreover, a term describing the random tunnelling of electrons between these two edge channels with amplitude $\xi(x)$ is included. These processes are due to the presence of impurities along the edge and their role in the dynamics will be discussed in the following. 

Looking more in detail into the form of the bosonic fields $\hat{\phi}_{1/2}(x)$ one has that they can be decomposed as 
\begin{equation}
\hat{\phi}_{1/2}(x)=\hat{\phi}^{(p)}_{1/2}(x)+\hat{\phi}^{(0)}_{1/2}(x).
\label{phi_decomposition}
\end{equation}
The first term in Eq.~(\ref{phi_decomposition}) is a plasmonic contribution
\begin{equation}
\hat{\phi}^{(p)}_{1/2}(x)=\sum^{+\infty}_{k=0}\sqrt{\frac{2\pi |\nu_{1/2}|}{k L}} e^{-\frac{k \alpha}{2}}
\left(\hat{b}_{1/2, k} e^{i k x} +\hat{b}^{\dagger}_{1/2, k} e^{-i k x}\right)
\label{plasmonic}
\end{equation}
which obeys the periodic boundary condition $\hat{\phi}^{(p)}_{1/2}(x+L)=\hat{\phi}^{(p)}_{1/2}(x)$. In Eq.~(\ref{plasmonic}) we have introduced the bosonic annihilation (creation) operator $\hat{b}_{1/2,k}$ ($\hat{b}^{\dagger}_{1/2,k}$), $\alpha$ is a short distance cut-off and where the wave-numbers are quantized according to
\begin{equation}
k=\frac{2 \pi}{L} m,
\end{equation}
with $m\in \mathbb{N}^{*}$.
 
The second term in Eq.~(\ref{phi_decomposition}) is a zero mode contribution
\begin{equation}
\hat{\phi}^{(0)}_{1/2}(x)= \pm \frac{2 \pi}{L}\hat{N}_{1/2}x-\nu_{1/2}\hat{\chi}_{1/2}
\end{equation}
where the operators $\hat{N}_{1/2}$ count the excess number of electron in the corresponding channel with respect to a given reference and $\hat{\chi}_{1/2}$ is an additional operator satisfying 
\begin{equation}
\left[\hat{\chi}_{j}, \hat{N}_{k} \right]=i \delta_{j,k}
\end{equation}
and such that $e^{\pm i \hat{\chi}_{1/2}}$ changes $\hat{N}_{1/2}$ by $\pm 1$ playing the role of a Klein factor.

Combining the operators $\hat{\phi}_{1/2}$, it is possible to define the new charged ($\hat{\phi}_{\rho}(x)$) and neutral ($\hat{\phi}_{\sigma}(x)$) fields given by 
\begin{eqnarray}
\hat{\phi}_{\rho}&=&\sqrt{\frac{3}{2}}\left(\hat{\phi}_{1}+\hat{\phi}_{2}\right)\\
\hat{\phi}_{\sigma}&=&\frac{1}{\sqrt{2}}\left(\hat{\phi}_{1}+3\hat{\phi}_{2}\right), 
\end{eqnarray}
with
\begin{equation}
\left[\phi_{\rho/\sigma}(x), \phi_{\rho/\sigma}(y) \right]=\pm i \pi \mathrm{sgn}(x-y).
\end{equation}
According to this, one can rewrite the Hamiltonian in Eq.~(\ref{H_D_12}) as
\begin{eqnarray}
\hat{H}_{D}&=&\frac{1}{4 \pi}\int^{L}_{0} dx \left[v_{\rho} \left(\partial_{x} \hat{\phi}_{\rho} \right)^{2}+v_{\sigma} \left(\partial_{x} \hat{\phi}_{\sigma} \right)^{2}+v_{\rho \sigma }\partial \hat{\phi}_{\rho} \partial \hat{\phi}_{\sigma}\right]\nonumber\\
&+& \int^{L}_{0} dx \left[\xi(x) e^{i \sqrt{2}\hat{\phi}_{\sigma}}+H.c. \right],
\label{H_D}
\end{eqnarray}
with 
\begin{eqnarray}
v_{\rho}&=& \frac{3}{2} v_{1}+\frac{1}{2} v_{2}-v_{12}\\
v_{\sigma}&=& \frac{1}{2} v_{1}+\frac{3}{2}v_{2}-v_{12}\\
v_{\rho \sigma}&=& -\sqrt{3}\left(v_{1}+v_{2}\right)+\left(\sqrt{3}+\frac{1}{\sqrt{3}} \right) v_{12}.
\end{eqnarray}
The very remarkable fact discussed in Refs.~\cite{Kane1994, Kane1995}, is that in presence of strong enough random electron tunneling $\xi(x)$ the system flows towards a fixed point with effective Hamiltonian description 
\begin{equation}
\hat{H}_{D}\approx\frac{1}{4 \pi}\int^{L}_{0} dx \left[v_{\rho} \left(\partial_{x} \hat{\phi}_{\rho} \right)^{2}+v_{\sigma} \left(\partial_{x} \hat{\phi}_{\sigma} \right)^{2}\right]
\label{H_D2}
\end{equation}
with $v_{\rho}\gg v_{\sigma}$. 

This so called \emph{disorder dominated} Hamiltonian will be the basis of following analysis. It is worth to note that, in a situation where it is possible to relax the condition of extremely precise conductance quantization, the same Hamiltonian in Eq.~(\ref{H_D}) can be derived  simply assuming strong Coulomb interaction between the modes and neglecting the random electron tunneling between the channels~\cite{Wen1995}. This strengthen the generality of the following analysis.

Taking into account again the decomposition into plasmonic and zero mode part, the quantum dot Hamiltonian can be rewritten as 

\begin{equation}
\hat{H}_{D}\approx \sum^{+\infty}_{m=1} m \left(E_{\rho} \hat{b}^{\dagger}_{ \rho, m}\hat{b}_{ \rho, m}+E_{\sigma}\hat{b}^{\dagger}_{ \sigma, m}\hat{b}_{\sigma, m}  \right)+\frac{3}{4}E_{\rho}\left(\hat{N}_{1}-\hat{N}_{2}\right)^{2}+\frac{1}{4}E_{\sigma}\left(\hat{N}_{1}-3\hat{N}_{2}\right)^{2}
\label{H_D_modes}
\end{equation}
where we have introduced the new bosonic annihilation (creation) operators $\hat{b}_{\rho/\sigma,m}$ ($\hat{b}^{\dagger}_{\rho/\sigma,m}$) and defined 
\begin{equation}
E_{\rho/ \sigma}=\frac{2 \pi}{L} v_{\rho/\sigma}.
\end{equation}

\subsubsection{Tunneling Hamiltonian}

We consider now the possibility of electron tunneling from the leads to the quantum dot and viceversa. Considering only the more relevant tunneling process, it is possible  to write the local tunneling Hamiltonian

\begin{equation}
\hat{H}_{T}=\sum_{l=L, R} \left[t_{l} \hat{\Psi}^{\dagger}_{l}(x_{l}) \hat{\Psi}_{D}(x_{l})+t^{*}_{l} \hat{\Psi}^{\dagger}_{D}(x_{l}) \hat{\Psi}_{l}(x_{l}) \right]
\end{equation}
with 
\begin{equation}
\hat{\Psi}_{D}(x)=\frac{1}{\sqrt{2 \pi \alpha}} e^{i \frac{\pi}{L} x}e^{i \hat{\phi}_{1}(x)}=\frac{1}{\sqrt{2 \pi \alpha}} e^{i \frac{\pi}{L} x}e^{i \left[\sqrt{\frac{3}{2}}\hat{\phi}_{\rho}(x)-\sqrt{\frac{1}{2}}\hat{\phi}_{\sigma}(x) \right]}
\end{equation}
the bosonized expression for the annihilation operator on an electron in the $\nu_{1}$ channel~\cite{Wen1995}, $t_{l}$ the (energy independent) tunneling amplitudes and $x_{L/R}$ the points where the local tunneling occurs. Finally the additional phase factor has been introduced in order to provide the proper boundary conditions for the electron field operator on a closed geometry~\cite{Geller1997, Braggio2006, Merlo2007}.  

\subsubsection{Hamiltonian of the systems}
According to the above discussions the complete Hamiltonian of the system can be written as 

\begin{equation}
\hat{H}=\hat{H}_{L}+\hat{H}_{R}+\hat{H}_{D}+\hat{H}_{T}.
\end{equation}
This will be used in the following to evaluate the tunneling rates. 


\subsection{Tunneling rates}

Starting from the Hamiltonian in Eq.~(\ref{H_D_modes}), a generic state of the quantum dot can be completely characterized in terms of the quantum numbers
\begin{equation}
|N_{1}, N_{2},\left\{n\right\}_{\rho}, \left\{n\right\}_{\sigma}\rangle
\end{equation}
with $N_{j}$ ($j=1,2$) the number of electron in the $j$-th channel and

\begin{equation}
\left\{n\right\}_{\rho/\sigma}=\left\{n_{\rho/\sigma,1}; n_{ \rho/\sigma,2}; \dots; n_{ \rho/\sigma, m}; \dots \right\}
\end{equation}
the occupation numbers of the charged and neutral plasmon modes for every wave number $m$. Within the sequential tunneling regime, where for each tunneling process only one electron is transferred from the leads to the dot and viceversa due to the action of $\hat{H}_{T}$, eventually exciting some plasmons, these quantum numbers are enough the completely describe the transport properties of the system. Moreover, due to the form of $\hat{H}_{T}$, tunneling events can only change $N_{1}$ leaving $N_{2}$ unaffected. 

Here, we assume a tunneling amplitude weak enough that the tunneling conductance remains smaller with respect to the conductance quantum $e^{2}/h$. Within this approximation we can consider the rate for the transition between an arbitrary initial state of the total system $|i\rangle$ to a final state $|f\rangle$ induced by the tunneling Hamiltonian $\hat{H}_{T}$. It can be evaluated using the Fermi's golden rule 

\begin{equation}
\Gamma(|i \rangle \rightarrow |f\rangle)= 2 \pi |\langle i |\hat{H}_{T}|f \rangle|^{2}
\delta (E_{f}-E_{i})
\end{equation}
with $E_{i}$ ($E_{f}$) the initial (final) energy. 

Integrating over the states of the electrons in the leads and taking into account the fact that they are characterized by the thermal distribution of free fermions in Eq.~(\ref{f_l}), one has that the total tunneling rates for adding or removing an electron from the channel at $\nu_{1}$ of the dot through the $l=L, R$ contact can be written respectively as~\cite{Kim2005, Kim2006, Frigeri2020} 

\begin{eqnarray}
&&\Gamma_{l}\left(N_{1}, N_{2}, \left\{n\right\}_{\rho}, \left\{n\right\}_{\sigma}\rightarrow N_{1}+1, N_{2}, \left\{n'\right\}_{\rho}, \left\{n'\right\}_{\sigma}\right)\nonumber\\
&=&\gamma_{l} \mathcal{M}\left(\left\{n\right\}_{\rho}, \left\{n'\right\}_{\rho} \right)\mathcal{M}\left(\left\{n\right\}_{\sigma}, \left\{n'\right\}_{\sigma} \right)\nonumber\\
&\times& f_{l}\left[\Delta_{+}\left(N_{1}, N_{2}, \left\{n\right\}_{\rho}, \left\{n\right\}_{\sigma}, \left\{n'\right\}_{\rho}, \left\{n'\right\}_{\sigma} \right) \right]
\label{Gamma_plus}
\end{eqnarray}

and

\begin{eqnarray}
&&\Gamma_{l}\left(N_{1}, N_{2}, \left\{n\right\}_{\rho}, \left\{n\right\}_{\sigma}\rightarrow N_{1}-1, N_{2}, \left\{n'\right\}_{\rho}, \left\{n'\right\}_{\sigma}\right)\nonumber\\
&=&\gamma_{l} \mathcal{M}\left(\left\{n\right\}_{\rho}, \left\{n'\right\}_{\rho} \right)\mathcal{M}\left(\left\{n\right\}_{\sigma}, \left\{n'\right\}_{\sigma} \right) \nonumber\\
&\times&\left\{1-f_{l}\left[-\Delta_{-}\left(N_{1}, N_{2}, \left\{n\right\}_{\rho}, \left\{n\right\}_{\sigma}, \left\{n'\right\}_{\rho}, \left\{n'\right\}_{\sigma} \right) \right]\right\}.
\label{Gamma_minus}
\end{eqnarray}

In the above equations we have defined 
\begin{equation}
\gamma_{l}=\frac{|t_{l}|^{2}}{vL}
\end{equation}

and the matrix elements taking into account the possible change of the excitation number in the plasmon sectors are given by~\cite{Frigeri2020}

\begin{equation}
\mathcal{M}\left(\left\{n\right\}, \left\{n'\right\} \right)=\frac{\prod^{m_{max}}_{m=1} e^{-\frac{1}{2m}}\left( \frac{1}{2m}\right)^{|n_{m}-n'_{m}|}\frac{n^{(<)}_{m}!}{n^{(>)}_{m}!}\left[\mathcal{L}^{|n_{m}-n'_{m}|}_{n^{(<)}_{m}}\left(\frac{1}{2m} \right)\right]^{2}}{\prod^{m_{max}}_{m=1} e^{-\frac{1}{2m}}}
\end{equation}
where 
\begin{eqnarray}
n^{(>)}_{m}&=&\mathrm{max}\left(n_{m}, n'_{m}\right)\\
n^{(<)}_{m}&=&\mathrm{min}\left(n_{m}, n'_{m}\right), 
\end{eqnarray}
$\mathcal{L}^{b}_{a}(x)$ are the associated Laguerre polynomials, and $m_{max}$ is an index such that $n_{m}=n'_{m}=0$ for every $m>m_{max}$. Notice that, in the following, we will consider the possibility of having different values of $m_{max}$ for charged and neutral sectors. This will allow us to consider tunneling events able to excite only the neutral sector, leaving the charged one unaffected. In the following we will demonstrate how this assumption is crucial in order to justify the experimentally observed enhancement of the Fano factor.

In Eqs.~(\ref{Gamma_plus}-\ref{Gamma_minus}) we have also introduced the short notation 

\begin{eqnarray}
&&\Delta_{\pm}\left(N_{1}, N_{2}, \left\{n\right\}_{\rho}, \left\{n\right\}_{\sigma}, \left\{n'\right\}_{\rho}, \left\{n'\right\}_{\sigma} \right)\nonumber\\
&=&E\left(N_{1}\pm1, N_{2}, \left\{n'\right\}_{\rho}, \left\{n'\right\}_{\sigma} \right)-E\left(N_{1}, N_{2}, \left\{n\right\}_{\rho}, \left\{n\right\}_{\sigma} \right)
\end{eqnarray}

with 

\begin{eqnarray}
&&E\left(N_{1}, N_{2}, \left\{n\right\}_{\rho}, \left\{n\right\}_{\sigma} \right)\nonumber\\
&=&\frac{3}{4}E_{\rho}\left(N_{1}-N_{2} \right)^{2}
+\frac{1}{4}E_{\sigma}\left(N_{1}-3N_{2} \right)^{2}+\sum^{+\infty}_{m=1} m \left(E_{\rho}n_{m, \rho}+E_{\sigma}n_{m, \sigma} \right)
\end{eqnarray}
in agreement with Eq.~(\ref{H_D_modes}).

In the following, we will focus on a regime where only a limited number of tunneling rates provide a relevant contribution to the transport, the other being suppressed either by the plasmon contribution or by the energy mismatch in the Fermi distributions. In particular, we will consider a three-state model elaborated in Refs.~\cite{Kim2005, Kim2006, Cavaliere2004} where an interaction-induced enhancement of the noise, and consequently of the Fano factor, has been reported.

\subsubsection{Three-state model}


Considering, as stated above, the symmetric voltage configurations $V_{L/R}=\pm V/2$ one can write the relevant tunneling rates 

\begin{eqnarray}
\mathcal{A}_{-}&=& \Gamma_{L}\left(0,0, \left\{0\right\}_{\rho},\left\{0\right\}_{\sigma}\rightarrow 1, 0, \left\{0\right\}_{\rho}, \left\{0\right\}_{\sigma}  \right)/\gamma_{L}\nonumber\\
&=& \frac{1}{1+\exp\left(\frac{3E_{\rho}+E_{\sigma}}{4 E_{T}}-\frac{E_{V}^\ast}{2 E_{T}} \right)}\\
\mathcal{A}_{+}&=& \Gamma_{R}\left(1,0, \left\{0\right\}_{\rho},\left\{0\right\}_{\sigma}\rightarrow 0, 0, \left\{0\right\}_{\rho}, \left\{0\right\}_{\sigma}  \right)/\gamma_{R}\nonumber\\
&=&\frac{\exp\left(\frac{3E_{\rho}+E_{\sigma}}{4 E_{T}}+\frac{E_{V}^\ast}{2E_{T}} \right)}{1+\exp\left(\frac{3E_{\rho}+E_{\sigma}}{4 E_{T}}+\frac{E_{V}^\ast}{2E_{T}} \right)}\\
\mathcal{B}_{-}&=& \Gamma_{L}\left(0,0, \left\{0\right\}_{\rho},\left\{1\right\}_{\sigma}\rightarrow 1, 0, \left\{0\right\}_{\rho}, \left\{0\right\}_{\sigma}  \right)/\gamma_{L}\nonumber\\
&=&\frac{1}{2}\frac{1}{1+\exp\left(\frac{3E_{\rho}}{4E_{T}}-\frac{3E_{\sigma}}{4E_{T}}-\frac{E_{V}^\ast}{2E_{T}} \right)}\\
\mathcal{B}_{+}&=& \Gamma_{R}\left(1,0, \left\{0\right\}_{\rho},\left\{0\right\}_{\sigma}\rightarrow 0, 0, \left\{0\right\}_{\rho}, \left\{1\right\}_{\sigma}  \right)/\gamma_{R}\nonumber\\
&=&\frac{1}{2}\frac{\exp\left(\frac{3E_{\rho}}{4E_{T}}-\frac{3E_{\sigma}}{4E_{T}}+\frac{E_{V}^\ast}{2E_{T}} \right)}{1+\exp\left(\frac{3E_{\rho}}{4E_{T}}-\frac{3E_{\sigma}}{4E_{T}}+\frac{E_{V}^\ast}{2E_{T}} \right)}.
\end{eqnarray}
In the above expressions we have introduced the short notations $E_{V}^\ast=eV$ and $E_{T}=k_{B}T$. Here, it is possible to highlight the role played by an unrelaxed dipolar plasmon leading to the emergence of different current-carrying processes. 

Within this picture the Fano factor, namely the ratio between the current fluctuation (noise) and the current itself, can be written as~\cite{Kim2005, Kim2006, Cavaliere2004} 

\begin{equation}
F=1+ 2 \left[\frac{\mathcal{A}_{+}\mathcal{B}_{+}\left(\mathcal{A}_{-}-\mathcal{B}_{-} \right)^{2}-\mathcal{A}_{-}\mathcal{B}_{-}\left(\mathcal{A}_{-}\mathcal{B}_{+}+\mathcal{A}_{+}\mathcal{B}_{-} \right)}{(\mathcal{A}_{+}\mathcal{B}_{-}+\mathcal{A}_{-}\mathcal{B}_{+}+\mathcal{A}_{-}\mathcal{B}_{-})^2} \right].
\end{equation}

This quantity is plotted in main text Fig.~4a) versus $E_{\sigma}$ and $E_\rho$, showing a robust enhancement of the Fano factor ($1\leq F\leq 2$) as a consequence of the interaction-induced separation into a fast propagating charged mode and a slow (counter)-propagating neutral mode in a range of parameters consistent with the experimental conditions.

\section{Details of the Thomas-Fermi Calculation}

This section describes the details of our Thomas-Fermi (TF) calculation, which predicts the quantum Hall droplet reconstruction at the QPC under experimentally relevant gate geometries. We closely follow the approach developed in Ref.~\cite{Cohen2025,Cohen2023} and briefly review it below. The total energy of the two-dimensional electron gas is written as a functional of the electron density $ n(\mathbf{r}) $,
\begin{equation}
\mathcal{E}[n(\mathbf{r})]
=
\frac{e^2}{2}
\int_{\mathbf{r}_1,\mathbf{r}_2}
n(\mathbf{r}_1)
V(\mathbf{r}_1,\mathbf{r}_2)
n(\mathbf{r}_2)
-
e
\int_{\mathbf{r}}
\Phi(\mathbf{r})
+
\mathcal{E}_{\rm xc}[n(\mathbf{r})].
\end{equation}
Here, the first term represents electron-electron (Hartree) interactions, the second term arises from gate-induced potentials, and $\mathcal{E}_{\rm xc}[n(\mathbf{r})]$ incorporates exchange-correlation. We adopt the local density approximation (LDA) by writing
\begin{equation}
\mathcal{E}_{\rm xc}[n(\mathbf{r})]
=
\int_{\mathbf{r}}
E_{\rm xc}(n(\mathbf{r})),\quad
E_{\rm xc}(n)
=
\int_{0}^{n}
dn'
\mu(n'),
\end{equation}
where $\mu(n)$ comes from our previously measured chemical potential in monolayer graphene at high magnetic field~\cite{Yang2021}. The minimization of $\mathcal{E}[n(\mathbf{r})]$ follows Ref.~\cite{Cohen2025,Cohen2023}, with a coarse-grained treatment of $n(\mathbf{r})$ at magnetic length $\ell_B$ scale.

A central new ingredient, not included in Ref.~\cite{Cohen2025,Cohen2023}, is the dielectric function $\epsilon_{\nu}(\mathbf{q})$ that encodes inter-Landau-level (LL) mixing at the level of random phase approximation (RPA). We implement
\begin{equation}
\epsilon_{\nu}(\mathbf{q})
=
1
-
V_{0}(\mathbf{q})
\Pi_{\nu}(\mathbf{q},\omega=0),
\end{equation}
where $V_{0}(\mathbf{q})$ is the bare Coulomb interaction, and $\Pi_{\nu}(\mathbf{q},\omega=0)$ sums over allowed LL transitions. The overall scale of $\Pi_{\nu}(\mathbf{q},\omega=0)$ depends on the fine-structure constant $\alpha_G$ in graphene. Previous studies place $\alpha_G$ in the range $(1.75,2.2)$ \cite{Shizuya2007,Yang2021}, and here we set $\alpha=2$. This modification removes the need to artificially boost $E_{\rm xc}$ by a factor of 1--2 to reproduce experiments in previous works \cite{Cohen2025,Cohen2023}.

We simulate a system at $B = 13$ T (magnetic length $\ell_B \approx 7$ nm), with a top gate at $d_{\rm t} = 60$ nm, a back gate at $d_{\rm b} = 30$ nm, and channel widths $w_{\rm EW} = w_{\rm NS} = 30$ nm. The hBN between gates and sample has anisotropic dielectric constants $\epsilon_{\parallel} = 6.6$ and $\epsilon_{\perp} = 3$. We discretize a $60\,\ell_B \times 60\,\ell_B$ region into grid spacing $\Delta x = \Delta y = \ell_B/5$. In momentum space, the gate-screened Coulomb kernel is
\begin{equation}
V(\mathbf{q})
=
\frac{e^2}{4\pi \epsilon_0}
\frac{
4\pi
\sinh(\beta d_{\rm t}|\mathbf{q}|)
\sinh(\beta d_{\rm b}|\mathbf{q}|)
}{
\sinh[\beta(d_{\rm t}+d_{\rm b})|\mathbf{q}|]
|\mathbf{q}|
}
\frac{
|F_{N}(\mathbf{q})|^2
}{
\epsilon_{\rm hBN}
\epsilon_{\nu}(\mathbf{q})
},
\end{equation}
where $\beta = \sqrt{\epsilon_{\parallel}/\epsilon_{\perp}}$, $\epsilon_{\rm hBN} = \sqrt{\epsilon_{\parallel} \epsilon_{\perp}}$, and $F_N(\mathbf{q})$ the usual LL form factor. For the zero energy LL in graphene, $F_0(\mathbf{q}) = e^{-\tfrac{1}{4}|\mathbf{q}|^2 \ell_B^2}$.
We further tune the gate voltages to fix the filling factor $\nu_{\mathrm{EW}} = 2$ beneath the east and west gates and $\nu_{\mathrm{NS}} = 0$ beneath the north and south gates, which follows experiment and creates a QPC region with $1 \lesssim \nu \lesssim 2$. We find both the filling factors $\nu_{\mathrm{NS}} = 0$ and $\Pi_{\nu}(\mathbf{q},\omega=0)$ crucial to reproduce the reconstructed islands anticipated at the QPC.

\section{Reconstruction at the QPC}

In Ref.~\cite{Cohen2025}, a Thomas-Fermi calculation similar to the one presented here demonstrated that a fractional island with filling $\nu = 1/3$ can appear at the saddle point of a QPC, provided the confining potential is sufficiently smooth compared to the Coulomb energy. In the present work, which focuses on a different bulk filling configuration ($\nu_{\mathrm{EW}} = 2$ and $\nu_{\mathrm{NS}} = 0$), we find that a similar reconstruction takes place at the QPC but instead yields an isolated quantum dot at $\nu = 2/3$. This observation underpins the noise model discussed in the main text.

The reconstruction is governed by the competition between two energy scales: the Coulomb energy $E_{C}$ and the confining-potential scale $E_{V}$. Concretely,
\begin{equation}
E_{C} = \frac{e^2}{4 \pi \epsilon_0 \epsilon_{\mathrm{hBN}}\,\ell_B} 
\approx 47.4 \ \mathrm{meV}, 
\quad
E_{V} = e\left(\frac{\partial \Phi}{\partial r}\right)\ell_B.
\end{equation}
Here, $E_{C}$ sets the characteristic Coulomb strength in the system, while $E_{V}$ measures how quickly the electrostatic potential varies across the magnetic length $\ell_B$. Generally, a smaller ratio $E_{V}/E_{C}$ favors the formation of reconstructed islands. By tuning gate voltages $V_{\mathrm{EW}},\,V_{\mathrm{NS}},\,V_{\mathrm{B}}$, we access $\,E_{V}/E_{C}\in[0.57,\,0.61]$.

\begin{figure}[h]
\centering
\includegraphics[width=.9\textwidth]{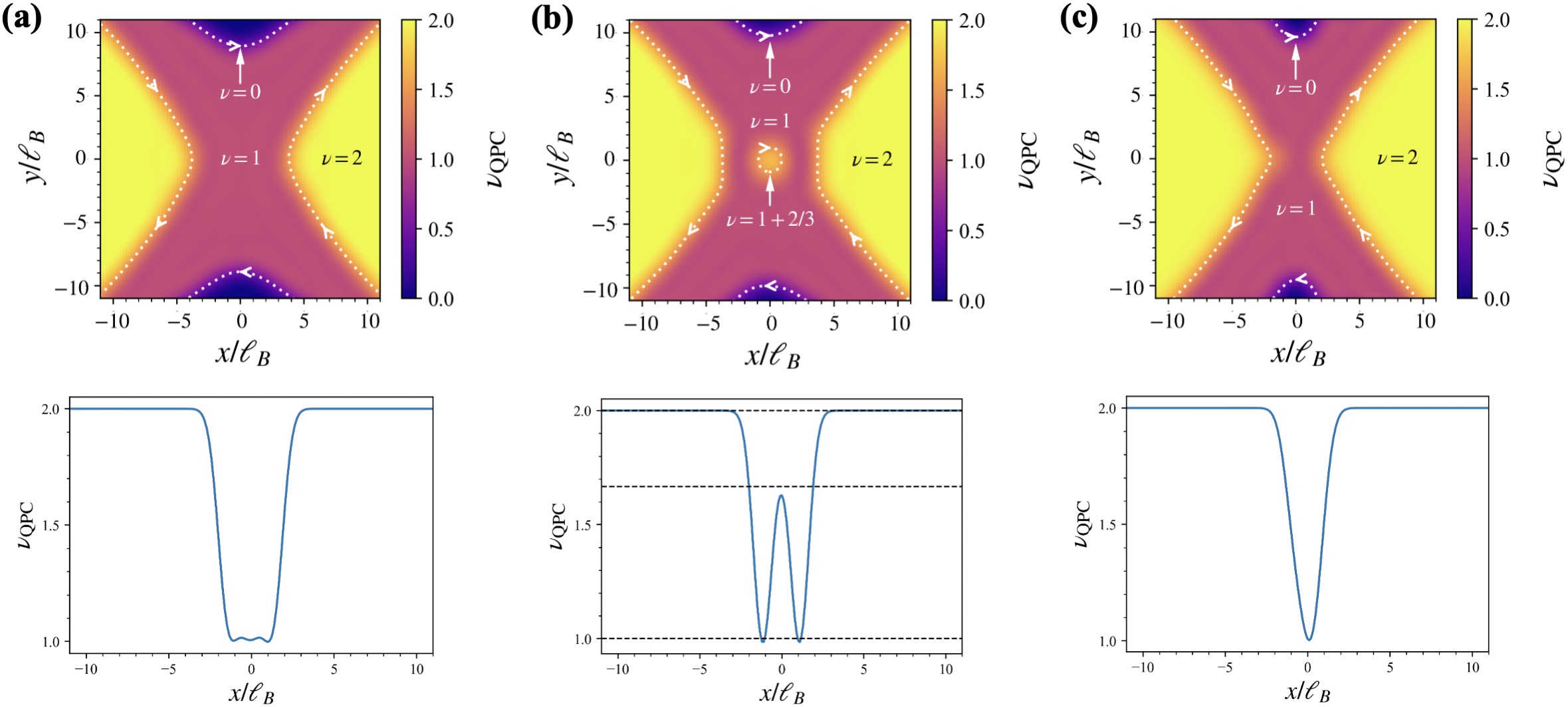}
\caption{\small
Reconstructed two-dimensional electron density profiles in a QPC geometry, along with one-dimensional cuts in the $x$-direction, for different values of $E_{V}/E_{C}$. 
(a) At $E_{V}/E_{C}=0.57$, there is a relatively wide $\nu=1$ channel beneath the E/W gates ($\nu_{\mathrm{EW}}=2$), with no localized dot. 
(b) Increasing to $E_{V}/E_{C}=0.59$ produces an island at the QPC center with $\nu=2/3$ (above the $\nu=1$ plateau). 
(c) Further increasing to $E_{V}/E_{C}=0.61$ narrows the $\nu=1$ channel until the fractional dot disappears.
}
\label{fig:S2}
\end{figure}

Figure~\ref{fig:S2} illustrates our main findings. We keep $\nu_{\mathrm{EW}}=2$ under the EW gates and $\nu_{\mathrm{NS}}=0$ under the NS gates, while varying $E_{V}/E_{C}$. When $E_{V}/E_{C}=0.57$, the $\nu=1$ channel is wide, and no localized state forms above $\nu=1$. As $E_{V}/E_{C}$ increases to $0.59$, a fractional island appears at $\nu=2/3$. This differs from the $\nu=1/3$ droplet in Ref.~\cite{Cohen2025} due to the changed bulk filling ($\nu_{\mathrm{NS}}=0$). Finally, at $E_{V}/E_{C}=0.61$, the channel narrows enough that the $\nu=2/3$ island vanishes, leaving a simpler edge structure.

In addition to requiring a suitable $E_{V}/E_{C}$ ratio, the formation of the $\nu=2/3$ island also depends on the width of the $\nu=1$ channel. If this channel is too narrow for a given $E_{V}/E_{C}$, the island cannot emerge; conversely, if it is too wide, the potential gradient is insufficient to confine electrons above $\nu=1$. Consequently, both the slope of the confining potential and the $\nu=1$ channel width work together to determine whether the fractional droplet materializes in the QPC region.

The fractional $\nu=2/3$ island encloses two $\nu=1/3$ edge channels, which can recombine into a charge mode and a neutral mode upon hybridization. This neutral mode could mediate tunneling between the $\nu=2$ bulk edges and the island's charge mode, and therefore enhance current fluctuations at the QPC. Consequently, the emergence of this fractional droplet is consistent with the observed noise signatures in our experiment.

\section{Semiclassical model}

The code of the semiclassical model is shown below. It consists in building charge transfer statistics over a large number of tunneling attempts, with a simple rule: whenever a tunneling attempt (parametrized in the code by the transmission probability $t$) is successful, the next attempt automatically succeeds. After this dual tunneling process the transmission probability is reset back to $t$. We compute the reflected current as the average number of failed attempts and its noise as the variance, yielding the Fano factor calculation shown in main text Fig.~3 and Supplemental Fig.~\ref{FigSup-Fanovstau-4configs}.

\newpage
\includepdf{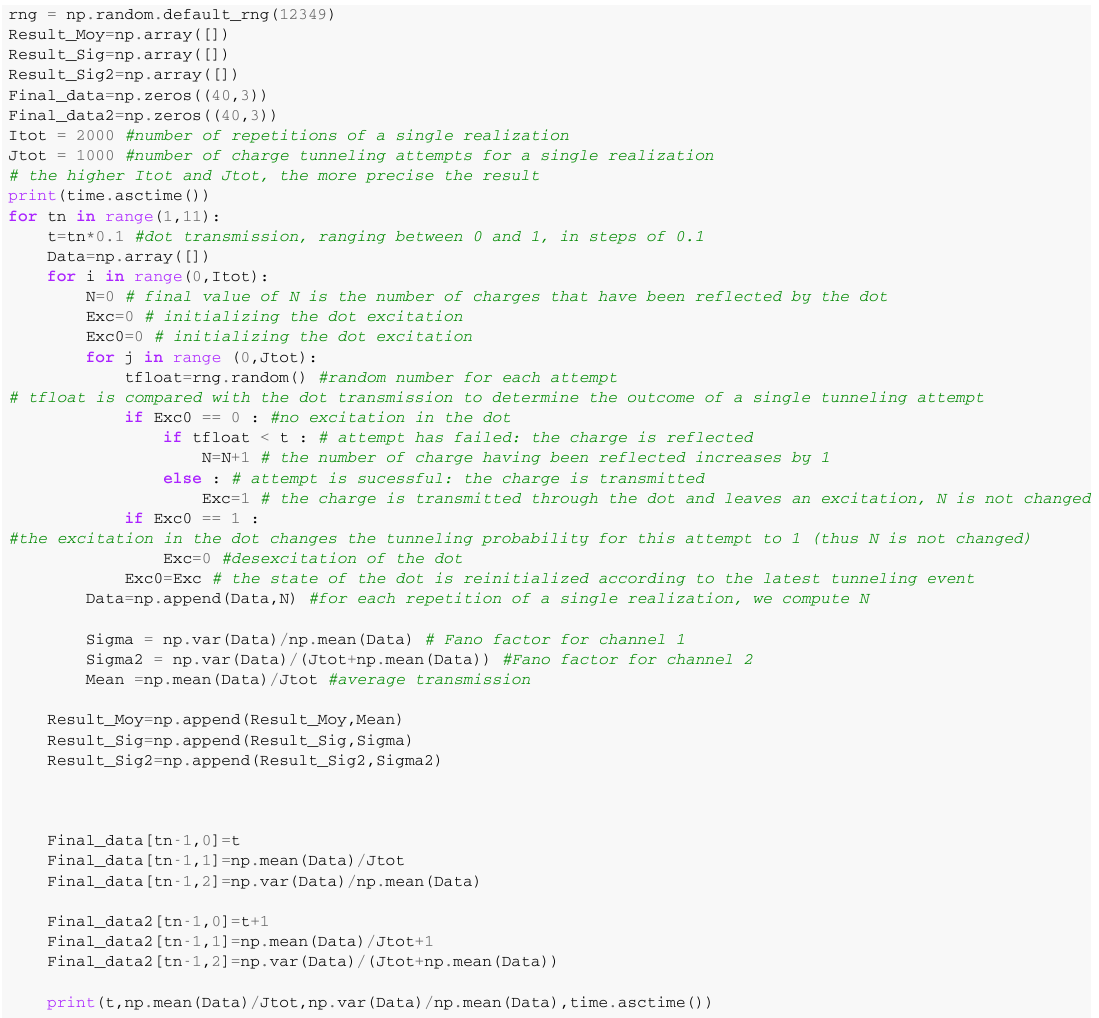}